\documentclass[twocolumn]{aastex631}
\usepackage{amsmath}

\newcommand{\hi}{\mbox{H{\small I}}}

\shorttitle{Star Formation in Green Valley Galaxies}
\shortauthors{Villanueva et al.}
\graphicspath{{./}{figures/}}

\begin{document}

\title{The EDGE-CALIFA survey: Molecular Gas and Star Formation Activity Across the Green Valley}

\author[0000-0002-5877-379X]{Vicente Villanueva}
\altaffiliation{ALMA-ANID Postdoctoral Fellow}
\affiliation{Department of Astronomy, University of Maryland, College Park, MD 20742, USA}
\affiliation{Departamento de Astronom\'ia, Universidad de Concepci\'on, Barrio Universitario, Concepci\'on, Chile}

\author[0000-0002-5480-5686]{Alberto D. Bolatto} 
\affiliation{Department of Astronomy, University of Maryland, College Park, MD 20742, USA}

\author[0000-0002-8765-7915]{Stuart N. Vogel}
\affiliation{Department of Astronomy, University of Maryland, College Park, MD 20742, USA}

\author[0000-0002-7759-0585]{Tony Wong}
\affiliation{Department of Astronomy, University of Illinois, Urbana, IL 61801, USA}

\author[0000-0002-2545-1700]{Adam K. Leroy}
\affiliation{Department of Astronomy, The Ohio State University, Columbus, OH 43210, USA}
\affiliation{Center for Cosmology and Astroparticle Physics, 191 West Woodruff Avenue, Columbus, OH 43210, USA}

\author[0000-0001-6444-9307]{Sebastian F. S\'anchez}
\affiliation{Instituto de Astronom\'ia, Universidad Nacional Aut\'onoma de M\'exico, A.P. 70-264, 04510 M\'exico, D.F., M\'exico}

\author[0000-0003-2508-2586]{Rebecca C. Levy}
\altaffiliation{NSF Astronomy and Astrophysics Postdoctoral Fellow}
\affiliation{Steward Observatory, University of Arizona, Tucson, AZ 85721, USA}

\author[0000-0002-5204-2259]{Erik Rosolowsky}
\affiliation{Department of Physics, University of Alberta, 4-181 CCIS, Edmonton, AB T6G 2E1, Canada}

\author[0000-0001-6498-2945]{Dario Colombo}
\affiliation{Argelander-Institut f\"ur Astronomie, Universit\"at Bonn, Auf dem H\"ugel 71, 53121 Bonn, Germany}

\author[0000-0002-2262-5875]{Veselina Kalinova}
\affiliation{Max Planck Institute for Radioastronomy, Auf dem H\"ugel 69, D-53121, Bonn, Germany}

\author[0000-0002-9511-1330]{Serena Cronin}
\affiliation{Department of Astronomy, University of Maryland, College Park, MD 20742, USA}

\author[0000-0003-1774-3436]{Peter Teuben}
\affiliation{Department of Astronomy, University of Maryland, College Park, MD 20742, USA}







\author{M\'onica Rubio}
\affiliation{Departamento de Astronom\'ia, Universidad de Chile, Casilla 36-D Santiago, Chile}

\author{Zein Bazzi}
\affiliation{Argelander-Institut f\"ur Astronomie, Universit\"at Bonn, Auf dem H\"ugel 71, 53121 Bonn, Germany}







\correspondingauthor{Vicente Villanueva}
\email{vvillanu@umd.edu}



\begin{abstract}

We present a $^{12}$CO($J$=2-1) survey of 60 local galaxies using data from the Atacama Large Millimeter/submillimeter Compact Array as part of the Extragalactic Database for Galaxy Evolution: the ACA EDGE survey. These galaxies all have integral field spectroscopy from the CALIFA survey \citep{Sanchez2012,Sanchez2016}. {Compared to other local galaxy surveys, ACA EDGE is designed to mitigate selection effects based on CO brightness and morphological type.} Of the 60 galaxies in ACA EDGE,   36 are on the star-formation main sequence, 13 are on the red sequence, and 11 lie in the ``green valley" transition between these sequences. We test how star formation quenching processes affect the star formation rate (SFR) per unit molecular gas mass, SFE$_{\rm mol}=$SFR/$M_{\rm mol}$, and related quantities in galaxies with stellar masses $10\leq$log[$M_\star/$M$_\odot$]$\leq11.5$ covering the full range of morphological types. We observe a systematic decrease of the molecular-to-stellar mass fraction ($R^{\rm mol}_{\star}$) with decreasing level of star formation activity, with green valley galaxies having also lower SFE$_{\rm mol}$ than galaxies on the main sequence. On average, we find that the spatially resolved SFE$_{\rm mol}$ within the bulge region of green valley galaxies is lower than in the bulges of main sequence galaxies if we adopt a constant CO-to-H$_2$ conversion factor, $\alpha_{\rm CO}$. While efficiencies in main sequence galaxies remain almost constant with galactocentric radius, in green valley galaxies we note a systematic increase of SFE$_{\rm mol}$, $R^{\rm mol}_{\star}$, and specific star formation rate, sSFR, with increasing radius. As shown in previous studies, our results suggest that although gas depletion (or removal) seems to be the most important driver of the star-formation quenching in galaxies transiting through the green valley, a reduction in star formation efficiency is also required during this stage.

\end{abstract}

\date{December 2023. Accepted for publication in ApJ}


\section{Introduction} \label{sec:intro}

Star formation activity plays a key role in driving the growth and evolution of galaxies. The production of stars is quantified through the star formation rate (SFR) which is in principle a function of the physical conditions in the dense interstellar medium (ISM). In the last decades, several studies have revealed a tight correlation for many galaxies between the integrated SFR and the stellar mass ($M_{\star}$) in galaxies, the so-called star-formation main sequence \citep[SFMS; e.g.,][]{Brinchmann2004,Daddi2007,Whitaker2012,Saintonge2016,Colombo2020}. This implies a useful galaxy classification in terms of their star-formation status: ``blue cloud'' galaxies, which show a direct correlation between $M_{\star}$ and SFR for active star-forming galaxies; ``red cloud,'' where galaxies exhibit low SFRs and no $M_{\star}$-SFR correlation; and the ``green valley'' (or transition galaxies; \citealt{Salim2007}). The bimodality of the SFMS suggests fundamental questions regarding the physical processes behind the transition from the SFMS through the green valley the red cloud, which is mostly linked to the cessation of star formation activity.

The term ``quenching''  has been adopted to include the variety of mechanisms behind the cessation of star formation activity in galaxies. In particular, \cite{Peng2010} suggest two different routes to classify quenching processes: ``environmental quenching'', which is coupled to the local environmental conditions that may drive the decrease (or cessation) of SFR; and ``mass quenching'', which refers to internal/intrinsic galaxy mechanisms affecting star formation. While environmental processes mostly take place in galaxies residing in high-density environments (e.g., galaxy clusters), encompassing a broad variety of environmental mechanisms (e.g., strangulation/starvation, \citealt{Larson1980,Balogh&Morris2000}; ram pressure stripping, \citealt{Gunn&Gott1972}, galaxy interactions, \citealt{Moore1996,Smith2010}), intrinsic mechanisms are usually associated with the activation and regulation of the physical processes driving star formation activity. Intrinsic quenching mechanisms are also expected to act differently depending on the structural components within galaxies, resulting in variations in the SFR when comparing bulges, bars, or disks. These intrinsic mechanisms have been broadly associated with fast quenching processes ($\lesssim100$ Myr; e.g.  \citealt{Bluck2020,Bluck2020b}), or slow ageing ($\sim0.5$--$1$ Gyr; e.g. \citealt{Corcho-Caballero2023}), which act in different ways to alter the physical conditions of the gas and span from strangulation (i.e, star formation continues until the reservoirs of cold gas are depleted; e.g., \citealt{Kawata&Mulchaey2008,Peng2015}) gas removal, either due to active galactic nuclei (AGN) suppression \citep[e.g.,][]{Oppenheimer2010, Page2012}, or via stellar feedback (e.g., SNe winds; \citealt{Oppenheimer2010}). 


Recent theoretical models have shown that some of these intrinsic mechanisms rely on modifying the physical properties of the ISM, thereby changing the efficiency by which the molecular gas is transformed into stars. \cite{Martig2009} proposed ``morphological quenching'', a process in which star formation is suppressed by the formation of a stellar spheroid. According to \cite{Martig2009}, morphological quenching reflects the stabilization of the disk by the dominant presence of a pressure-supported stellar spheroid, which replaces the stellar disk. The stabilization of the gas is a consequence of two effects: i) the steep potential well induced by the spheroid, and ii) the increase of a stellar spheroid relative to the stellar disk suppresses the growth of perturbations in the gaseous disk. This process provides a mechanism through which early-type galaxies (ETGs) lose their ability to form stars even in the presence of significant cold gas reservoirs (e.g., \citealt{Martig2013}). Gravitational instability is key to increase the SFR. In a simple model, stability is typically estimated by the Toomre $Q$ parameter, $Q=\frac{\kappa \sigma}{\epsilon G \Sigma}$ \citep[][]{Toomre1964}, where $\sigma$ is the one-dimensional dispersion velocity of the gas, $\sigma_{\rm gas}$, and $\Sigma$ is the surface density of an infinitelly thin disk; $\kappa$ is the epicyclic frequency, which is linked to {the steepness} of the gravitational potential, and is of order the angular velocity $\Omega$. Axisymmetric instabilities, which create rings that break-up into clouds, can grow in the disks if $Q<1$. \cite{Martig2009} suggested that morphological quenching is the severe suppression of star-formation activity in a massive gaseous disk when it is embedded in a dominant bulge that stabilizes the gas (i.e., resulting in $Q\gg1$). When compared with the star-formation activity in spirals, the difference in disk stability in ETGs arises from two main effects: i) the high central concentration of the stellar mass in ETGs increases $\kappa$, consequently increasing the tidal forces as well; and ii) the spheroidal distribution of stars dilutes the self-gravity of the gas, and therefore gravitational collapse cannot counteract the tidal forces, preventing the assembly of star-forming clumps. Also through numerical simulations, \cite{Gensior2020} show that spheroids drive turbulence and increase $\sigma_{\rm gas}$, increase the virial parameter, and cause the turbulent pressure to increase towards a galaxy center; all these are mostly dependant on the bulge mass ($M_{\rm b}$). They also find that turbulence increases for more compact and more massive bulges. 
\noindent Although morphological quenching is a process able to reduce the star formation during a well-defined time range of a galaxy lifetime 
\noindent (t$\approx 7-11$ Gyr; \citealt{Martig2009}), it is still not clear to what degree the ageing in ETGs is driven by this mechanism, the reduction of the molecular gas content, or a combination of multiple processes.

By obtaining high-resolution CO data, the new generation of mm/submm telescopes has allowed us to analyze in detail how the physical conditions of the molecular gas vary between the different structural components within galaxies in the local Universe. In addition, multiwavelength galaxy surveys have revealed the interplay between the different components of the ISM and their role behind star formation activity. 
\noindent In this work, we present the ALMA large mm/submm Compact Array Extragalactic Database for Galaxy Evolution, the ACA EDGE survey. We investigate the star-formation activity in 60 nearby massive galaxies using Atacama Compact Array (ACA) observations of the CO(2-1) emission line in combination with optical Integrated Field Unit (IFU) data from the CALIFA survey \citep[][]{Sanchez2012}.

This paper is organized as follows: Section \ref{S2_Observations} presents the main features of the ACA EDGE survey, including the sample selection, data processing, and the ancillary data. In Section \ref{S3_Methods} we explain the methods applied to analyze the data and the equations used to derive the physical quantities. Finally, in Section \ref{S4_Results} we present our results and discussion, and in Section \ref{S5_Conclusions} we summarize the main conclusions. Throughout this work, we assume a $\Lambda$CDM cosmology, adopting the values $\Omega_{\Lambda}=0.7$, $\Omega_{\rm DM}=0.3$ and H$_{\circ}$=$69.7$ km s$^{-1}$ Mpc$^{-1}$.

\section{OBSERVATIONS}
\label{S2_Observations}
\begin{figure}
\hspace{-0.175cm}
\includegraphics[width=8.75cm]{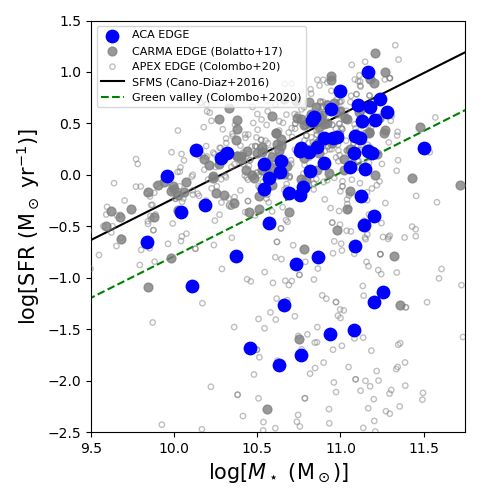}
\caption{SFR-$M_\star$ relation for the 60 galaxies in the ACA EDGE survey (blue circles), sampling the whole range of $z=0$ galaxy behavior for log$[M_*/$M$_{\odot}]\approx 10-11.5$, including the star formation main sequence and quenched systems below it. Gray filled and unfilled circles are CARMA EDGE and APEX EDGE galaxies included in \cite{Bolatto2017} and \cite{Colombo2020}, respectively. The black-solid and dashed-green lines correspond to the best-linear fit for star-formation main sequence (\citealt{Cano-Diaz2016}) and green valley (\citealt{Colombo2020}) galaxies, respectively. ACA EDGE galaxies constitute a sample of the local universe with good statistical characteristics and are easy to volume-correct to characterize the star formation activity in nearby massive galaxies.} 
\label{fig_1}
\end{figure}

\begin{figure*}[h!]
\centering

\hspace{-0.5cm}
\begin{tabular}{c}

\includegraphics[width=0.85\textwidth]{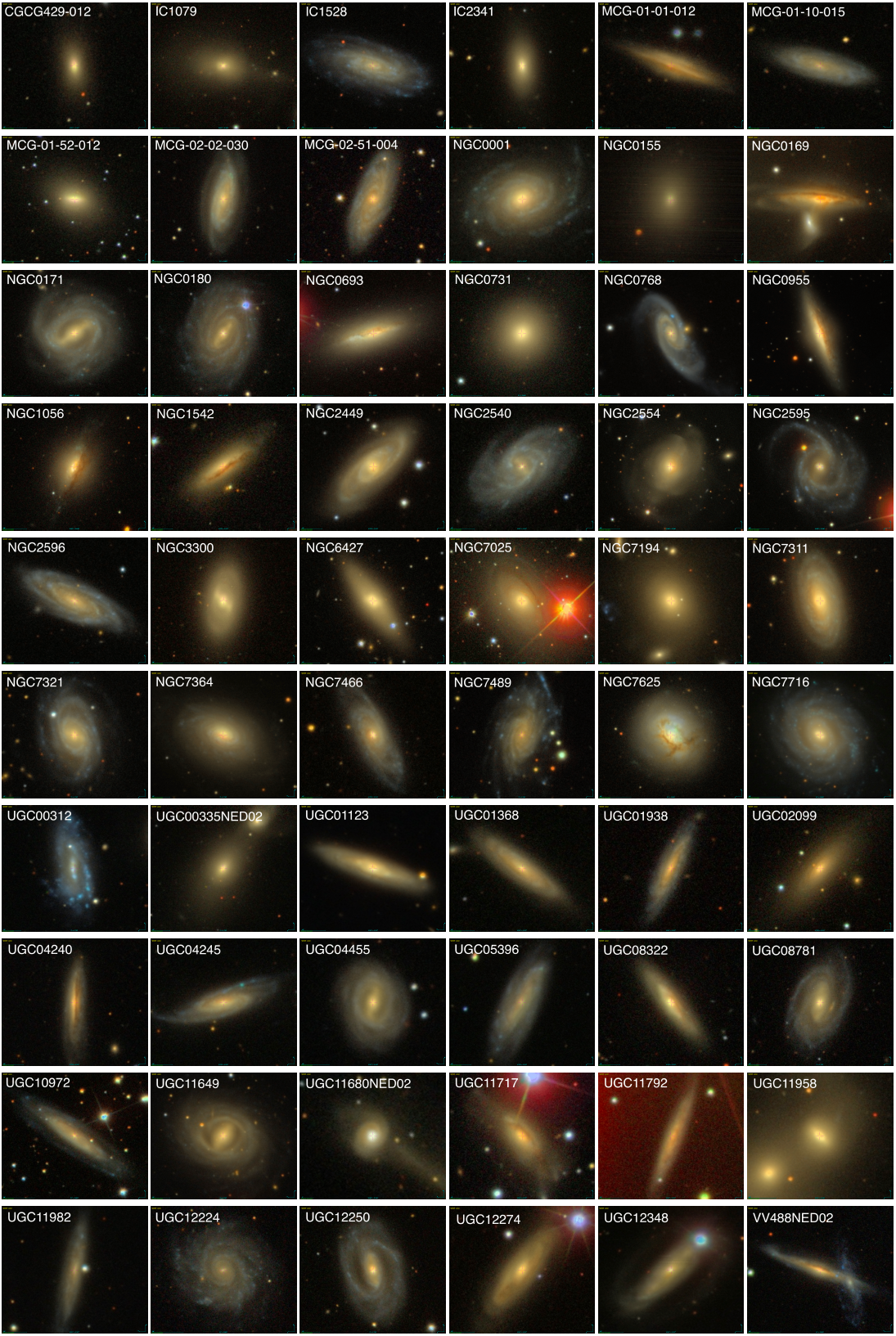}

\end{tabular}
\caption{SDSS $r$ (red channel), $i$ (green channel), and $z$-bands (blue channel) composite images for the 60 galaxies encompassed by the ACA EDGE survey. These local galaxies show a broad variety of morphologies representative of the distribution of galaxies in the local universe. This enables one of the main ACA EDGE goals, to analyze the star-formation quenching mechanisms at different evolutionary stages.}
\label{fig_2}
\end{figure*}

\subsection{The ACA EDGE sample}
\label{sample_selection}

We used the ALMA large mm/submm Compact Array (ACA) to observe 60 galaxies drawn from the third public data release of the Calar Alto Legacy Integral Field Area survey Data Release 3 \citep[][]{Sanchez2016b}, in the context of the Extragalactic Database for Galaxy Evolution (EDGE) surveys. {Previous CO surveys focus mainly (or exclusively) on ``main sequence'' or star forming galaxies selected either due to their SFR/$M_\star$, morphology, or IR brightness (e.g., the HERA CO Line Extragalactic Survey, HERACLES, \citealt{Leroy2008,Leroy2013}; the Herschel Reference Survey, HRS, \citealt{Boselli2010}; the James Clerk Maxwell Telescope Nearby Galaxies Legacy Survey, NGLS, \citealt{Wilson2012}; the CO Legacy Database for GALEX Arecibo SDSS Survey, COLD GASS and the extended COLD GASS, xCOLD GASS, \citealt{Saintonge2011a,Saintonge2017}; The Extragalactic Database for Galaxy Evolution-Calar Alto Legacy Integral Field Area, the EDGE-CALIFA survey, \citealt{Bolatto2017}; Virgo Environment Traced in CO survey, VERTICO, \citealt{Brown2021}; the Physics at High Angular resolution in Nearby Galaxies-ALMA survey, PHANGS-ALMA, \citealt{Leroy2021}).  ACA EDGE was designed to probe into the low SFR/$M_\star$ regime to study processes associated with galaxy quenching.} CALIFA observed over 800 galaxies with IFU spectroscopy at Calar Alto selected from a combination of the Sloan Digital Sky Survey (SDSS; \citealt{York2000,Alam2015}) and an extension of galaxies that fulfilled the observational setup (see \citealt{Sanchez2016b} for more details), reflecting the $z=0$ galaxy population with log$[M_{\star}/$M$_\odot$]$=9-11.5$ in a statistically meaningful manner \citep[][]{Walcher2014}. ACA EDGE targets a subsample of CALIFA galaxies with declination appropriate to observe with ALMA ($\delta < 30^{\circ}$) and stellar mass $M_\star > 10^{10}$ M$_\odot$, so that CO can be readily detected and metallicity effects are not too severe. We impose no selection on SFR in order to cover the full range of star formation activities in this mass range and enable the study of quenching. 
The ACA EDGE survey complements the main science goals of the CARMA EDGE survey \citep[][ galaxies also drawn from CALIFA; see Fig. \ref{fig_1}]{Bolatto2017}, which encompasses CO observations for 126 CALIFA galaxies at $\sim {4.5}''$ resolution but with significant biases. Although CARMA EDGE-selected galaxies cover a broader range of masses (log$[M_\star/$M$_\odot$] $= 9.1–11.5$), it mostly focused on late-type, far-IR detected galaxies that are rich in molecular gas (hence actively star-forming), with morphological types mainly spanning from Sa to Scd. The ACA EDGE survey was designed to complement it by increasing the coverage of early-type galaxies, thus adding more red cloud galaxies to CARMA EDGE in order to drive more statistically significant results.
A total of 60 galaxies were observed in CO(2–1) by the ALMA Cycle 7 project 2019.2.00029.S (P.I. A. D. Bolatto). The galaxies are listed in Table \ref{table_1}; SDSS images are shown in Figure \ref{fig_2}. Optical inclinations and east-of-north position angles (PA) are taken from HyperLEDA\footnote{https://leda.univ-lyon1.fr/} and recomputed (when applicable) using fits files from SDSS $z$-band photometry (see \S \ref{Basic_equations}).

\begin{table*}
\hspace{-.75cm}
\resizebox{.93\linewidth}{!}{ 
\begin{tabular}{ccccccccccccc}
\hline\hline
Name & R.A. & Decl.  & $i$  & P.A. & Redshift & rms & $\theta_{\rm min}$ & $\theta_{\rm maj}$ & P.A.$_{\rm beam}$ & Distance & $R_{\rm e}$ & $r_{25}$ \\
  & (J2000) & (J2000) & (deg) & (deg) &  & (mK) & (\arcsec) & (\arcsec) & (deg) & (Mpc) & (\arcsec) & (\arcsec) \\
 (1) & (2) & (3) & (4) & (5) & (6) & (7) & (8) & (9) & (10) & (11) & (12) & (13) \\
\hline
CGCG429-012 & 22$^{\rm h}$36$^{\rm m}$49.8$^{\rm s}$ & 14$^{\circ}$23$\arcmin$13$.\arcsec$1   & 60 & 4* & 0.01751 & 12 & 5.3 & 9.7 & -36.0 & 84 & 5.3 & 25.0 \\  
IC1079 & 14$^{\rm h}$56$^{\rm m}$36.1$^{\rm s}$ & 09$^{\circ}$22$\arcmin$11$.\arcsec$1        & 50 & 81* & 0.02907 & 16 & 6.3 & 7.7 & -83.0 & 114 & 19.3 & 43.4 \\  
IC1528 & 00$^{\rm h}$05$^{\rm m}$05.3$^{\rm s}$ & 07$^{\circ}$05$\arcmin$36$.\arcsec$3        & 70 & 72* & 0.01250 & 13 & 4.6 & 8.2 & -77.0 & 48 & 16.9 & 64.1 \\  
IC2341 & 08$^{\rm h}$23$^{\rm m}$41.4$^{\rm s}$ & 21$^{\circ}$26$\arcmin$05$.\arcsec$5        & 60 & 2* & 0.01701 & 18 & 6.2 & 8.0 & -72.0 & 75 & 8.0 & 35.2 \\  
MCG-01-01-012 & 23$^{\rm h}$59$^{\rm m}$10.8$^{\rm s}$ & 04$^{\circ}$11$\arcmin$30$.\arcsec$6 & 90 & 70* & 0.01883 & 13 & 4.4 & 8.1 & 89.0 & 78 & 11.2 & 47.5 \\  
MCG-01-10-015 & 03$^{\rm h}$38$^{\rm m}$39.1$^{\rm s}$ & 05$^{\circ}$20$\arcmin$50$.\arcsec$4 & 73 & 75* & 0.01343 & 12 & 5.1 & 8.3 & -48.0 & 90 & 11.8 & 43.4 \\  
MCG-01-52-012 & 20$^{\rm h}$37$^{\rm m}$49.9$^{\rm s}$ & 06$^{\circ}$05$\arcmin$26$.\arcsec$7 & 43 & 72* & 0.01270 & 14 & 5.0 & 8.4 & -45.0 & 100 & 4.5 & 35.2 \\  
MCG-02-02-030 & 00$^{\rm h}$30$^{\rm m}$07.3$^{\rm s}$ & 11$^{\circ}$06$\arcmin$49$.\arcsec$1 & 70 & 170* & 0.01179 & 13 & 4.4 & 8.3 & -76.0 & 42 & 13.9 & 55.9 \\  
MCG-02-51-004 & 20$^{\rm h}$15$^{\rm m}$39.8$^{\rm s}$ & 13$^{\circ}$37$\arcmin$19$.\arcsec$3 & 68 & 159* & 0.01876 & 12 & 5.2 & 8.1 & -43.0 & 90 & 15.8 & 45.4 \\  
NGC0001 & 00$^{\rm h}$07$^{\rm m}$15.8$^{\rm s}$ & 27$^{\circ}$42$\arcmin$29$.\arcsec$7       & 34 & 110* & 0.01511 & 15 & 4.9 & 11.7 & -24.0 & 74 & 9.2 & 47.5 \\  
NGC0155 & 00$^{\rm h}$34$^{\rm m}$40.0$^{\rm s}$ & 10$^{\circ}$45$\arcmin$59$.\arcsec$4       & 40 & 169* & 0.02053 & 13 & 5.1 & 7.9 & -47.0 & 65 & 13.5 & 44.4 \\  
NGC0169 & 00$^{\rm h}$36$^{\rm m}$51.7$^{\rm s}$ & 23$^{\circ}$59$\arcmin$25$.\arcsec$3       & 72 & 87* & 0.01525 & 15 & 5.0 & 10.1 & -31.0 & 98 & 19.4 & 45.4 \\  
NGC0171 & 00$^{\rm h}$37$^{\rm m}$21.5$^{\rm s}$ & 19$^{\circ}$56$\arcmin$03$.\arcsec$3       & 33 & 101* & 0.01277 & 14 & 4.9 & 8.0 & -47.0 & 57 & 15.8 & 61.3 \\  
NGC0180 & 00$^{\rm h}$37$^{\rm m}$57.7$^{\rm s}$ & 08$^{\circ}$38$\arcmin$06$.\arcsec$7       & 45 & 163* & 0.01743 & 13 & 5.6 & 8.9 & -30.0 & 61 & 20.2 & 65.6 \\  
NGC0693 & 01$^{\rm h}$50$^{\rm m}$30.8$^{\rm s}$ & 06$^{\circ}$08$\arcmin$42$.\arcsec$8       & 90 & 106* & 0.00498 & 17 & 5.1 & 8.9 & -43.0 & 17 & 11.6 & 62.7 \\  
NGC0731 & 01$^{\rm h}$54$^{\rm m}$56.2$^{\rm s}$ & 09$^{\circ}$00$\arcmin$38$.\arcsec$9       & 20 & 155* & 0.01296 & 12 & 5.1 & 8.2 & -49.0 & 55 & 10.4 & 50.9 \\  
NGC0768 & 01$^{\rm h}$58$^{\rm m}$40.9$^{\rm s}$ & 00$^{\circ}$31$\arcmin$45$.\arcsec$2       & 68 & 28* & 0.02308 & 13 & 5.3 & 8.3 & -46.0 & 80 & 15.6 & 46.5 \\  
NGC0955 & 02$^{\rm h}$30$^{\rm m}$33.1$^{\rm s}$ & 01$^{\circ}$06$\arcmin$30$.\arcsec$3       & 90 & 19* & 0.00489 & 17 & 5.2 & 8.5 & -44.0 & 24 & 9.4 & 80.7 \\  
NGC1056 & 02$^{\rm h}$42$^{\rm m}$48.3$^{\rm s}$ & 28$^{\circ}$34$\arcmin$26$.\arcsec$8       & 53 & 162* & 0.00528 & 15 & 4.8 & 10.9 & -27.0 & 30 & 7.9 & 55.9 \\  
NGC1542 & 04$^{\rm h}$17$^{\rm m}$14.1$^{\rm s}$ & 04$^{\circ}$46$\arcmin$53$.\arcsec$9       & 90 & 127* & 0.01235 & 18 & 4.8 & 8.2 & -42.0 & 70 & 9.5 & 36.9 \\  
NGC2449 & 07$^{\rm h}$47$^{\rm m}$20.2$^{\rm s}$ & 26$^{\circ}$55$\arcmin$49$.\arcsec$1       & 70 & 135* & 0.01652 & 15 & 6.0 & 7.6 & -76.0 & 43 & 12.8 & 43.4 \\  
NGC2540 & 08$^{\rm h}$12$^{\rm m}$46.4$^{\rm s}$ & 26$^{\circ}$21$\arcmin$42$.\arcsec$6       & 55 & 123* & 0.02088 & 17 & 5.7 & 9.7 & -21.0 & 75 & 15.4 & 37.2 \\  
NGC2554 & 08$^{\rm h}$17$^{\rm m}$53.5$^{\rm s}$ & 23$^{\circ}$28$\arcmin$19$.\arcsec$9       & 47 & 153* & 0.01365 & 18 & 5.9 & 8.3 & -61.0 & 60 & 17.5 & 61.3 \\  
NGC2595 & 08$^{\rm h}$27$^{\rm m}$41.9$^{\rm s}$ & 21$^{\circ}$28$\arcmin$44$.\arcsec$7       & 35 & 30* & 0.01429 & 18 & 5.8 & 8.5 & -63.0 & 64 & 24.1 & 49.8 \\  
NGC2596 & 08$^{\rm h}$27$^{\rm m}$26.4$^{\rm s}$ & 17$^{\circ}$17$\arcmin$02$.\arcsec$3       & 68 & 63* & 0.01964 & 17 & 5.3 & 8.5 & -32.0 & 82 & 11.7 & 41.4 \\  
NGC3300 & 10$^{\rm h}$36$^{\rm m}$38.4$^{\rm s}$ & 14$^{\circ}$10$\arcmin$16$.\arcsec$1       & 57 & 173* & 0.01012 & 16 & 5.8 & 7.7 & -76.0 & 47 & 13.3 & 45.4 \\  
NGC6427 & 17$^{\rm h}$43$^{\rm m}$38.5$^{\rm s}$ & 25$^{\circ}$29$\arcmin$38$.\arcsec$1       & 70 & 35* & 0.01088 & 15 & 4.9 & 10.3 & -29.0 & 45 & 8.9 & 46.5 \\  
NGC7025 & 21$^{\rm h}$07$^{\rm m}$47.3$^{\rm s}$ & 16$^{\circ}$20$\arcmin$09$.\arcsec$1       & 54 & 44* & 0.01639 & 13 & 4.9 & 9.1 & -49.0 & 75 & 18.2 & 59.9 \\  
NGC7194 & 22$^{\rm h}$03$^{\rm m}$30.9$^{\rm s}$ & 12$^{\circ}$38$\arcmin$12$.\arcsec$4       & 43 & 17* & 0.02713 & 13 & 5.6 & 9.5 & -30.0 & 123 & 11.9 & 39.5 \\  
NGC7311 & 22$^{\rm h}$34$^{\rm m}$06.7$^{\rm s}$ & 05$^{\circ}$34$\arcmin$11$.\arcsec$6       & 62 & 11* & 0.01495 & 13 & 5.1 & 8.9 & -41.0 & 61 & 10.6 & 44.4 \\  
NGC7321 & 22$^{\rm h}$36$^{\rm m}$28.0$^{\rm s}$ & 21$^{\circ}$37$\arcmin$18$.\arcsec$5       & 56 & 17* & 0.02372 & 14 & 4.9 & 9.7 & -30.0 & 104 & 12.0 & 42.4 \\  
NGC7364 & 22$^{\rm h}$44$^{\rm m}$24.3$^{\rm s}$ & 00$^{\circ}$09$\arcmin$43$.\arcsec$5       & 54 & 66* & 0.01605 & 14 & 4.5 & 8.4 & 86.0 & 68 & 10.6 & 49.8 \\  
NGC7466 & 23$^{\rm h}$02$^{\rm m}$03.4$^{\rm s}$ & 27$^{\circ}$03$\arcmin$10$.\arcsec$1       & 66 & 25* & 0.02483 & 13 & 5.2 & 11.6 & -19.0 & 92 & 12.6 & 46.5 \\  
NGC7489 & 23$^{\rm h}$07$^{\rm m}$32.6$^{\rm s}$ & 22$^{\circ}$59$\arcmin$53$.\arcsec$6       & 63 & 165* & 0.02071 & 14 & 5.0 & 9.7 & -30.0 & 70 & 16.6 & 36.9 \\  
NGC7625 & 23$^{\rm h}$20$^{\rm m}$30.0$^{\rm s}$ & 17$^{\circ}$13$\arcmin$35$.\arcsec$0       & 40 & 45* & 0.00557 & 15 & 5.3 & 7.9 & -53.0 & 24 & 9.8 & 44.4 \\  
NGC7716 & 23$^{\rm h}$36$^{\rm m}$31.4$^{\rm s}$ & 00$^{\circ}$17$\arcmin$50$.\arcsec$2       & 44 & 34* & 0.00851 & 15 & 4.9 & 7.5 & -69.0 & 36 & 14.2 & 54.6 \\  
UGC00312 & 00$^{\rm h}$31$^{\rm m}$23.9$^{\rm s}$ & 08$^{\circ}$28$\arcmin$00$.\arcsec$6      & 63 & 7* & 0.01424 & 15 & 5.4 & 7.9 & -57.0 & 57 & 13.3 & 44.4 \\  
UGC00335NED02 & 00$^{\rm h}$33$^{\rm m}$57.3$^{\rm s}$ & 07$^{\circ}$16$\arcmin$05$.\arcsec$9 & 50 & 147* & 0.01812 & 13 & 5.4 & 9.0 & -31.0 & 78 & 16.6 & 44.4 \\  
UGC01123 & 01$^{\rm h}$34$^{\rm m}$07.9$^{\rm s}$ & 01$^{\circ}$01$\arcmin$56$.\arcsec$2      & 75 & 70* & 0.01615 & 12 & 5.2 & 8.4 & -43.0 & 54 & 9.8 & 36.9 \\  
UGC01368 & 01$^{\rm h}$54$^{\rm m}$13.1$^{\rm s}$ & 07$^{\circ}$53$\arcmin$01$.\arcsec$1      & 73 & 51* & 0.02653 & 12 & 5.5 & 7.0 & -70.0 & 108 & 10.9 & 40.5 \\
UGC01938 & 02$^{\rm h}$28$^{\rm m}$22.1$^{\rm s}$ & 23$^{\circ}$12$\arcmin$52$.\arcsec$7      & 78 & 155* & 0.02108 & 14 & 5.2 & 9.4 & -25.0 & 96 & 8.6 & 35.2 \\  
UGC02099 & 02$^{\rm h}$37$^{\rm m}$13.0$^{\rm s}$ & 21$^{\circ}$34$\arcmin$04$.\arcsec$0      & 66 & 138* & 0.02737 & 14 & 5.4 & 10.4 & -31.0 & 118 & 13.7 & 35.2 \\  
UGC04240 & 08$^{\rm h}$08$^{\rm m}$06.1$^{\rm s}$ & 14$^{\circ}$50$\arcmin$16$.\arcsec$3      & 76 & 178* & 0.02886 & 12 & 5.3 & 8.3 & 87.0 & 163 & 10.2 & 37.8 \\  
UGC04245 & 08$^{\rm h}$08$^{\rm m}$45.7$^{\rm s}$ & 18$^{\circ}$11$\arcmin$39$.\arcsec$0      & 70 & 107* & 0.01733 & 17 & 5.0 & 9.0 & -32.0 & 75 & 14.5 & 42.4 \\  
UGC04455 & 08$^{\rm h}$31$^{\rm m}$32.8$^{\rm s}$ & 01$^{\circ}$11$\arcmin$51$.\arcsec$8      & 47 & 13* & 0.03044 & 16 & 5.0 & 7.7 & -81.0 & 128 & 7.9 & 25.0 \\  
UGC05396 & 10$^{\rm h}$01$^{\rm m}$40.4$^{\rm s}$ & 10$^{\circ}$45$\arcmin$23$.\arcsec$0      & 75 & 145* & 0.01798 & 13 & 5.7 & 7.7 & -87.0 & 76 & 13.9 & 44.4 \\  
UGC08322 & 13$^{\rm h}$15$^{\rm m}$00.9$^{\rm s}$ & 12$^{\circ}$43$\arcmin$31$.\arcsec$0      & 73 & 36* & 0.02540 & 15 & 5.8 & 7.9 & -82.0 & 98 & 8.5 & 33.7 \\  
UGC08781 & 13$^{\rm h}$52$^{\rm m}$22.7$^{\rm s}$ & 21$^{\circ}$32$\arcmin$22$.\arcsec$0      & 50 & 161* & 0.02513 & 15 & 6.3 & 7.9 & -61.0 & 115 & 12.0 & 47.5 \\  
UGC10972 & 17$^{\rm h}$46$^{\rm m}$21.8$^{\rm s}$ & 26$^{\circ}$32$\arcmin$36$.\arcsec$9      & 78 & 55* & 0.01539 & 15 & 4.8 & 10.7 & -28.0 & 63 & 19.0 & 62.7 \\  
UGC11649 & 20$^{\rm h}$55$^{\rm m}$27.6$^{\rm s}$ & 01$^{\circ}$13$\arcmin$30$.\arcsec$9      & 42 & 90* & 0.01252 & 14 & 5.3 & 8.4 & -42.0 & 60 & 14.5 & 43.4 \\  
UGC11680NED02 & 21$^{\rm h}$07$^{\rm m}$45.8$^{\rm s}$ & 03$^{\circ}$52$\arcmin$40$.\arcsec$4 & 40 & 205* & 0.02615 & 12 & 4.7 & 8.3 & 89.0 & 111 & 9.9 & 20.8 \\  
UGC11717 & 21$^{\rm h}$18$^{\rm m}$35.4$^{\rm s}$ & 19$^{\circ}$43$\arcmin$07$.\arcsec$0      & 60 & 39* & 0.02088 & 19 & 5.5 & 8.6 & -54.0 & 90 & 11.8 & 36.1 \\  
UGC11792 & 21$^{\rm h}$42$^{\rm m}$12.7$^{\rm s}$ & 05$^{\circ}$36$\arcmin$55$.\arcsec$1      & 78 & 160* & 0.01586 & 18 & 5.1 & 8.2 & 87.0 & 68 & 17.4 & 40.5 \\  
UGC11958 & 22$^{\rm h}$14$^{\rm m}$46.8$^{\rm s}$ & 13$^{\circ}$50$\arcmin$27$.\arcsec$2      & 50 & 143* & 0.02618 & 13 & 5.5 & 9.8 & -31.0 & 112 & 17.0 & 59.9 \\  
UGC11982 & 22$^{\rm h}$18$^{\rm m}$52.9$^{\rm s}$ & 01$^{\circ}$03$\arcmin$31$.\arcsec$2      & 78 & 171* & 0.01554 & 14 & 4.5 & 8.3 & -87.0 & 69 & 12.3 & 36.9 \\  
UGC12224 & 22$^{\rm h}$52$^{\rm m}$38.3$^{\rm s}$ & 06$^{\circ}$05$\arcmin$37$.\arcsec$2      & 45 & 46* & 0.01156 & 13 & 5.2 & 8.9 & -41.0 & 50 & 22.4 & 44.4 \\  
UGC12250 & 22$^{\rm h}$55$^{\rm m}$35.8$^{\rm s}$ & 12$^{\circ}$47$\arcmin$24$.\arcsec$9      & 53 & 12* & 0.02405 & 12 & 5.4 & 9.4 & -36.0 & 92 & 12.8 & 49.8 \\  
UGC12274 & 22$^{\rm h}$58$^{\rm m}$19.5$^{\rm s}$ & 26$^{\circ}$03$\arcmin$43$.\arcsec$3      & 73 & 142* & 0.02538 & 13 & 5.0 & 11.4 & -23.0 & 110 & 12.4 & 40.5 \\  
UGC12348 & 23$^{\rm h}$05$^{\rm m}$18.8$^{\rm s}$ & 00$^{\circ}$11$\arcmin$22$.\arcsec$3      & 75 & 136* & 0.02510 & 13 & 5.1 & 7.1 & -74.0 & 108 & 12.4 & 38.6 \\  
VV488NED02 & 22$^{\rm h}$56$^{\rm m}$50.8$^{\rm s}$ & 08$^{\circ}$58$\arcmin$03$.\arcsec$1    & 82 & 73* & 0.01632 & 16 & 4.1 & 8.8 & -86.0 & 70 & 13.3 & 62.7 \\ 
\hline
\end{tabular}}
\caption{The ACA EDGE target sample. \textbf{\textbf Notes:} Column (1): galaxy name; (2) R.A. (J2000) of the galaxy optical center; (3) Decl. (J2000) of the galaxy optical center; (4) optical SDSS $r$-band inclination; (5) optical SDSS $r$-band position angle, calculated east of north. If no (*) is given, we give the receding side of the galaxy. (6) stellar redshift; (7) root mean square flux in 10 km s$^{-1}$ channels; (8) minor axis of the synthesized beam; (9) major axis of the synthesized beam; (10) position angle of the synthesized beam; (11) distance; (12) effective radius; (13) optical size of the major axis measured at 25 mag arcsec$^{-2}$ in the $B$ band.  Columns (2) and (3) are drawn from the NASA/IPAC Extragalactic Databse, NED. Column (4) is derived by finding the best fit for the SDSS $z$-band contours at $r\sim r_{25}$ as described in \S \ref{Basic_equations}. Columns (5) and (6) are drawn from HyperLEDA,
  except when kinematic information overruled. Column (11) is derived from column (6) assuming a $\Lambda$CDM cosmology with $\Omega_{\Lambda}=0.7$, $\Omega_{\rm DM}=0.3$, and H$_{\circ}$=$69.7$ km s$^{-1}$ Mpc$^{-1}$. Column (12) is taken from CALIFA. Column (13) is taken from \cite{deVaucouleurs1991}.}
\label{table_1}
\end{table*}

\subsection{The CO data}
\label{alma_data}

CO observations of our ACA-only project were taken between December 2019 and September 2021, spending between 15 and 43 minutes on-source for each galaxy. 
\noindent We set a spectral bandwidth of $\approx$ 1980 MHz and a raw spectral resolution of $\sim$ 1.938 MHz $\approx$ 2.5 km s$^{-1}$. The scheduling blocks were designed to detect the CO(2-1) emission line down to a root mean square (rms) spanning from $\sim$12 to 18 mK at 10 km s$^{-1}$ channel width (corresponding to mass surface density of $\sim0.9$ to $1.2$ M$_{\odot}$ pc$^{-2}$), and from $\sim$5\arcsec\ to 7\arcsec \ angular resolution (i.e., probing physical scales $\sim 1.5$ kpc at the distance of ACA EDGE galaxies), depending on the declination of the source.

{Each galaxy was observed in a Nyquist spaced mosaic (between 10 and 14 pointings) }
\noindent aligned with the major axis, covering the source out to $r_{25}$. As mentioned in the previous section, we obtained 7m ACA observations for 60, with an FoV$\sim$1.2\arcmin. Finally, we obtain 5$\sigma$ CO detections for 46 ACA EDGE galaxies, giving a detection rate of $\sim77$\%. 

\begin{figure*}
  \includegraphics[width=17.cm]{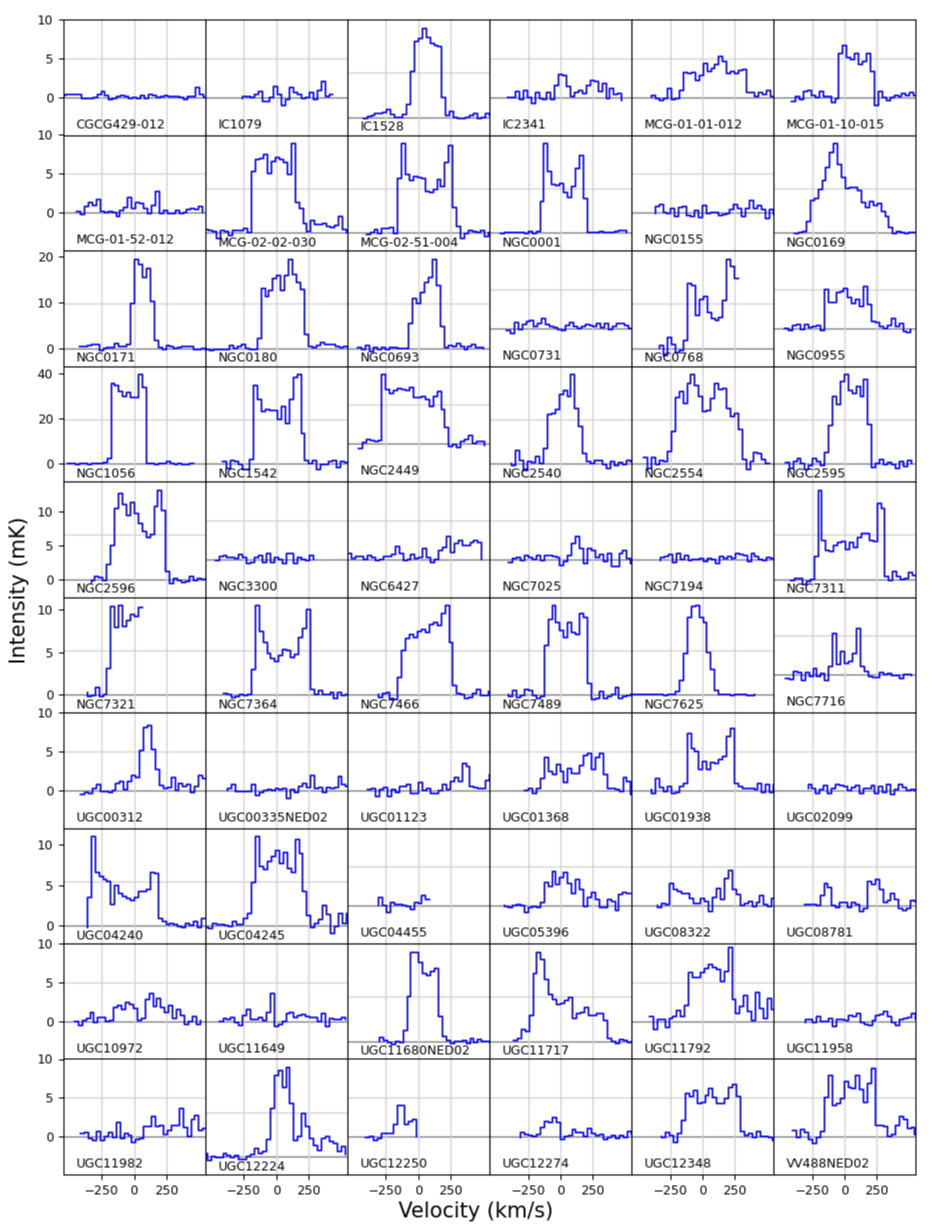}
  \caption{CO(2-1) spectra for ACA datacubes convolved to 1.1\arcmin \, and 30 km s$^{-1}$ channel width for the 60 galaxies. The spectra are taken from the central pixel located at the optical center (columns 2 and 3 in Table \ref{table_1}), and velocities are centered on the stellar redshift.}
  \label{fig_spectra}
\end{figure*}

\begin{figure*}
\hspace{.5cm}
  \includegraphics[width=16.cm]{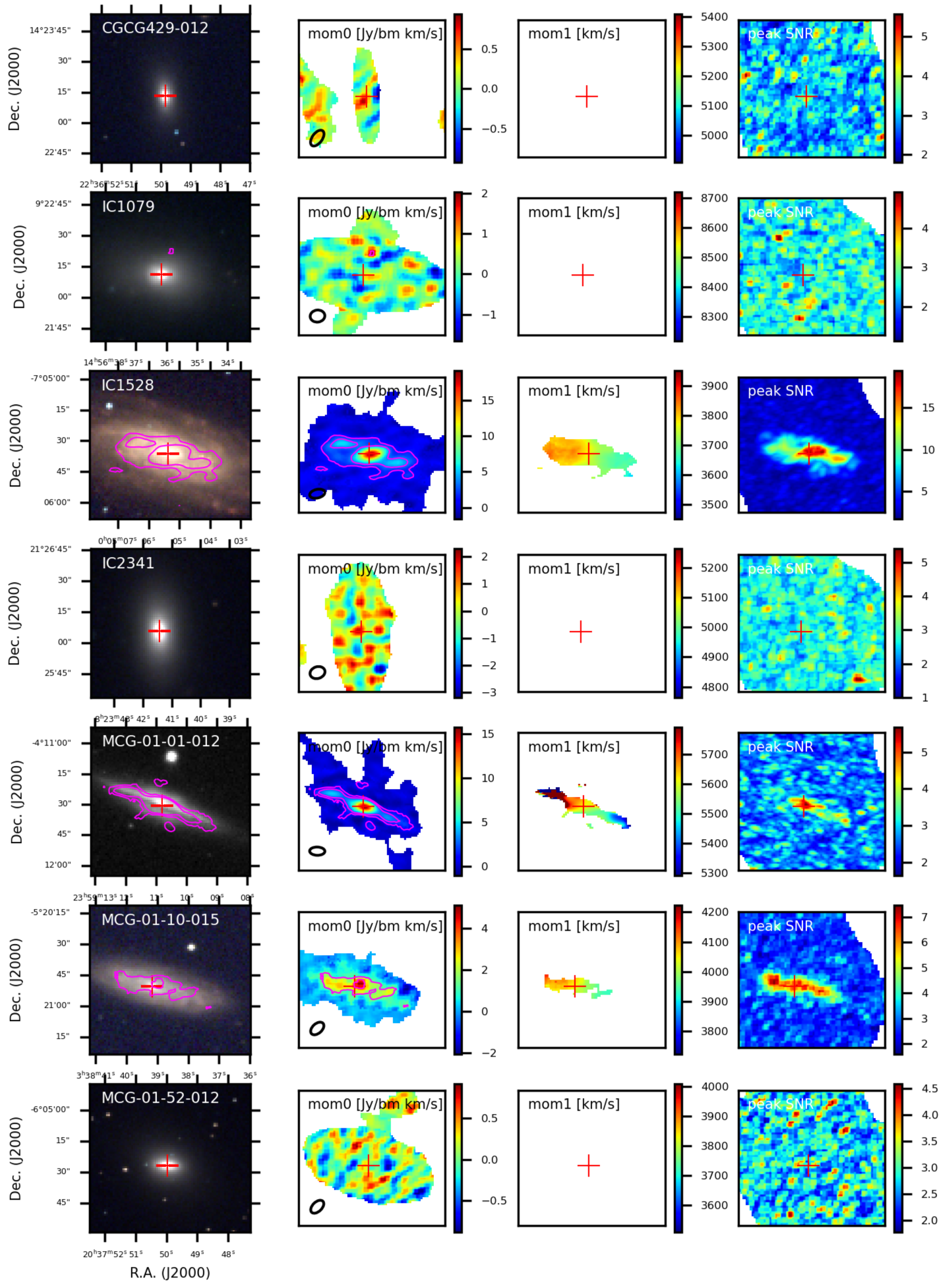}
  \caption{ACA EDGE data products for each galaxy. Panels cover an area of 1.25\arcmin $\times$ 1.25\arcmin. The first panel shows the SDSS $riz$ multicolor image with contours from our integrated intensity masked map overlaid. Contours correspond to $2\sigma$ and $5\sigma$ CO(2-1) emission line levels. From left to right, the following panels show the CO(2-1) emission line intensity (moment 0, in units of Jy/beam km s$^{-1}$), velocity (moment 1, in units of km s$^{-1}$), and signal-to-noise peak maps, respectively. The red crosses are the optical centers (columns 2 and 3 of Table \ref{table_1}). The black ellipses in the left bottom corner are the beam size of the CO(2-1) data. Panels for the remainder of the survey can be found in the Appendix.}
  \label{fig_mom0_1}
\end{figure*}

\subsection{Data reduction and products}
\label{data_processing}

We used $uv$ data delivered by ALMA and calibrated by the observatory pipeline \citep{Hunter2023}, then we imaged the CO$(J = 2–1)$ emission from each target using the PHANGS--ALMA Imaging Pipeline Version 2.1 \cite[][]{Leroy2021b}. Both the calibration and imaging utilized the Common Astronomy Software Applications ({\tt CASA}; \citealt{CASA2022}). The data were calibrated in {\tt CASA} 5.6.1-8 for data taken in 2019 and 2020 and {\tt CASA} 6.2.1-7 for data taken in 2021. We ran the PHANGS--ALMA imaging pipeline in CASA version 5.6.1-8.

Briefly, the PHANGS--ALMA imaging pipeline is designed to produce accurate images of extended spectral line emission. The pipeline combines all $uv$ data for a given target on a common spectral grid, subtracts continuum emission, and then carries out a multi-step deconvolution. This includes an initial multi-scale clean (we used scales of $0$, 5, and 10$\arcsec$) with a relatively high S/N$\approx 4$ threshold, followed by a single scale clean that uses an automatically generated, more restrictive clean mask and cleans down to S/N$\approx1$ by default. We used a Briggs weighting parameter of $=0.5$ \citep{Briggs1995} to achieve a good compromise between synthesized beam size and signal to noise. We used a channel width of 5.08 km s$^{-1}$ and adopted the local standard of rest (LSR) as our velocity reference frame, using the radio definition of velocity. After the initial imaging, the pipeline convolves the cube to have a round synthesized beam, converts the image to units of Kelvin, and downsamples the pixel gridding to save space while still Nyquist sampling the beam. See \citet{Leroy2021b} for more details. We did not use the noise modeling or product creation portions of the PHANGS--ALMA pipeline, but instead used software based on previous EDGE work.

All cubes were visually inspected for obvious problems or imaging errors. We note that NGC 0768, NGC 6427, NGC 7321, and UGC 12250 have incomplete CO line coverage since their emission peaks are located at the edge of the ACA spectral window. Although these galaxies have $5\sigma$ CO detections (see \S \ref{global_relations}), the CO line emission flux should just be taken as lower-limits.

We calculate moment maps and radial profiles using a masked cube. The construction of our mask follows a two-step procedure. We first create a mask using the CO cube, following the procedure for a ``dilated mask'' detailed in \citet{Bolatto2017}. This procedure includes in the mask areas around spectral peaks detected at $\gtrsim3.5\sigma$ significance (for more details, see \citealt{Rosolowsky&Leroy2006,Bolatto2017}). {We put together a second mask using information that is independent of the CO cube. We then use three different procedures to generate this mask, and choose the one that recovers the most CO emission flux. These procedures are: 
\begin{itemize}
    \item {\it H$\alpha$ mask (33 galaxies)}: We construct a mask using the central H$\alpha$ velocity map from CALIFA, and including around it a velocity region [-FWHM,+FWHM] following the FWHM prescription from Figure 2 in \citet{Villanueva2021}. H$\alpha$ spaxels with SNR$<5$ in intensity are excluded from the mask. This approach assumes that the kinematics of the CO are similar to the kinematics of the H$\alpha$ (e.g., \citealt{Levy2018}). 
    
   
    \item {\it Rotation mask (25 galaxies)}: We construct a mask assuming a very simple generic rotation curve that assumes the velocity is constant for $r>5\arcsec$ and increases linearly inside this radius. We adopt the maximum apparent rotation velocity reported in \hi\ by HyperLEDA  \citep[{\tt vmaxg} calculated from the 21-cm line, which we call $V_{\rm \hi,max}$ here;][]{Bottinelli1982,Bottinelli1990}, and adopt the systemic stellar velocity from CALIFA. We include the same velocity region around this central velocity as for the previous mask. This mask only extends to $r=r_{25}$. The direction of rotation is decided based on the H$\alpha$ or CO velocity field (if available) or ultimately if neither are detected based on comparing the flux recovered between the two senses of rotation. This approach assumes that the galaxy is predominantly rotating, and that the CO emission spans the same velocities as the \hi.
    
    
    \item {\it Flat mask (2 galaxies)}: We construct a mask centered at the stellar systemic velocity, including all the channels inside the maximum apparent rotation velocity reported by HyperLEDA and extending out to $r=0.5\,r_{25}$. This approach does not assume any particular kinematics and is the most relaxed of the three, although it will also include more noise.
\end{itemize}
Our final step is to combine (through a logical {\tt OR} operation) the best mask derived from this procedure  with the dilated mask obtained from the CO, in order to obtain the final mask. 

\begin{figure}
\hspace{-0.175cm}
\includegraphics[width=8.5cm]{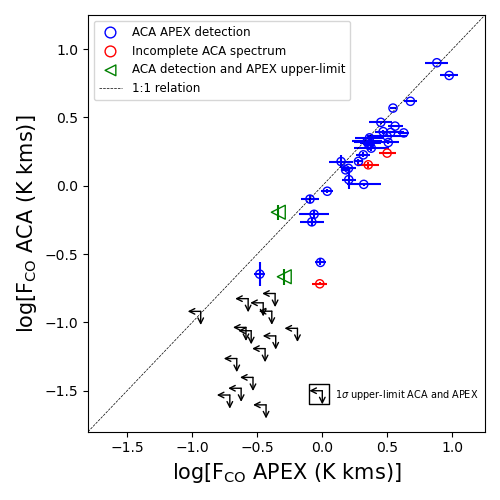}
\caption{Comparison of the integrated CO($J$=2-1) emission line flux between ACA (this work) and APEX \citep[][]{Colombo2020} datasets for 51 ACA galaxies. ACA fluxes are derived after convolving datacubes to match the APEX angular resolution (26.3\arcsec). The red dots correspond to NGC 0768, NGC 7321, and UGC 12250, which have incomplete ACA spectral coverage (see Fig. \ref{fig_spectra}). The green arrows are UGC 08322 and UGC 12274, which are detected by ACA but not APEX (see Table \ref{table_2}). The figure shows good agreement between ACA and APEX fluxes. However, fluxes measured by APEX are on average $\sim$20\% brighter than in ACA, likely due to calibration differences. Note that a lack of a detection by ACA in a 26\arcsec\ beam does not imply the source is not detected by ACA: for interferometric data convolution results in removing visibilities in long baselines (and thus collecting area and sensitivity).}
\label{fig_apex}
\end{figure}

We generate moment 0 maps (integrated intensity of the spectrum along the spectral axis) from the CO(2-1) spectral line cubes, in units of Jy/beam km s$^{-1}$, and after multiplying them by our mask (see Fig. \ref{fig_mom0_1}). To obtain the uncertainties of the moment 0 maps, we compute the rms in the signal-free part of the spectrum in each spaxel, $\sigma_{\rm i}$, and use the equation

\begin{equation}
u_{\rm i} =  \sigma_{\rm i} \sqrt{N} \Delta v.
\label{eq_uncertainties}
\end{equation}

\noindent Here, $N$ is the number of channels included by the mask and $\Delta v$ is the channel width (in km s$^{-1}$). We also compute the velocity (moment 1) and peak signal-to-noise ratio (SNR$_{\rm peak}$) maps. 

The moment 1 maps (or CO velocity maps, in units of km s$^{-1}$) are derived by multiplying the CO datacubes by the mask and using

\begin{equation}
M_{\rm 1,i} =  \frac{\Sigma I_{\rm i,j} v_{\rm j}}{M_{\rm 0,i}},
\label{eq_moment1}
\end{equation}

\noindent where $I_{\rm i,j}$ is the CO intensity in the $j$th spectral channel of the $i$th spaxel, $v_{\rm j}$ is the velocity of the $j$th channel, and $M_0$ is the moment 0 map. Finally, we blank the pixels outside the 2$\sigma$ contour for $M_0$. 
We also computed maps of the peak SNR, SNR$_{\rm peak,i}$, at each position.  We use the following equation:

\begin{equation}
{\rm SNR_{\rm peak}} =  \frac{max(I_{\rm i,j})}{\sigma_{\rm i}},
\label{eq_peak_snr}
\end{equation}
\noindent where $max(I_{\rm i,j})$ is the maximum value of the CO intensity within the spectrum of the $i$th spaxel. Both velocity and SNR$_{\rm peak}$ maps are included in Figure \ref{fig_mom0_1}}.

We compare the $^{12}$CO(2-1) integrated fluxes for ACA EDGE galaxies to those from \cite{Colombo2020}, 
\noindent who report $^{12}$CO(2-1) fluxes for 51 of our galaxies using APEX observations at 26.3\arcsec \ resolution and 30 km s$^{-1}$ channel width (as part of the APEX EDGE survey). The APEX EDGE survey arises from either the necessity of exploring whether star-formation quenching is driven by the reduction in molecular gas content, a change in the star formation efficiency of the molecular gas, or both. To address this, \cite{Colombo2020} use the $^{12}$CO(1–0) maps from the EDGE survey included in \cite{Bolatto2017} in combination with APEX 12CO(2-1) measurements. With these maps, they investigate the center of more than 470 galaxies selected from the CALIFA survey \citep[][]{Sanchez2012} at different quenching stages. To compare the fluxes, we convolve our CO datacubes to match the APEX angular resolution and we take the spectrum of the pixel located at the galaxy center, 
correcting by the recommended APEX main beam antenna efficiency (for the PI230 receiver at this frequency, $\eta_{\rm mb}=0.80$). Finally, we integrate the spectra over a spectral window defined by visual inspection (typically $\sim$500 km s$^{-1}$ wide). Uncertainties are computed by deriving the RMS from the signal-free part of the spectrum and using Equation \ref{eq_uncertainties}. For non-detections, we estimate 1$\sigma$ upper-limits by computing the RMS over the velocity window given by $V_{\rm \hi,max}$ and using Equation \ref{eq_uncertainties}. Discrepancies between both measurements can in principle be attributed to inconsistencies in calibration, flux that is resolved out or lost due to imperfect deconvolution for ACA measurements, or pointing for APEX. Although there are some discrepancies between the two datasets and a handful of cases with incomplete ACA spectral coverage, Figure \ref{fig_apex} shows that there is reasonable consistency between the ACA and APEX integrated CO fluxes. On average, we find that the median ACA-to-APEX flux ratio is 0.82. 

\subsection{The CALIFA survey and ancillary data}
\label{califa}

The Calar Alto Legacy Integral Field Area survey, CALIFA \citep[][]{Sanchez2012}, comprises a sample of over 800 galaxies at $z\approx0$. The data were acquired using the PMAS/PPAK IFU instrument \citep[][]{Roth2005} at the 3.5~m telescope of the Calar Alto Observatory. PMAS/PPAK uses 331 fibers each with a diameter of $2.7\arcsec$ in an hexagonal shape covering a FoV of a square arcminute. Its average spectral resolution is $\lambda / \Delta \lambda \sim 850$ at ${\rm \sim 5000 \AA}$ for a wavelength range that spans from $\lambda=3745$ to ${\rm 7300 \AA}$. As mentioned in \S \ref{sample_selection}, CALIFA galaxies are angular size-selected such that their isophotal diameters, $D_{25}$, match well with the PMAS/PPAK FoV. They range from $45$\arcsec\ to $80$\arcsec \ in the SDSS $r$-band \citep[][]{Walcher2014}. The CALIFA survey uses a data reduction pipeline designed to produce datacubes with more than $5000$ spectra with a sampling of $1\arcsec \times 1\arcsec$. For more details, see \cite{Sanchez2016b}. These cubes are processed using {\tt PIPE3D} \citep{Sanchez2015,Sanchez2016_2} to generate maps of derived quantities.

The final data compilation also contains ancillary data, including information from HyperLEDA, NASA/IPAC Extragalactic Databse (NED\footnote{https://ned.ipac.caltech.edu/}), among others.

\section{Methods and products}
\label{S3_Methods}

\subsection{Basic equations and assumptions}
\label{Basic_equations}

To compute the extinction-corrected SFRs, we estimate the extinction \citep[based on the Balmer decrement; see][]{Bolatto2017} for each 1\arcsec\ spaxel using the equation:

\begin{equation}
A_{\rm H{\alpha}} = 5.86 \log{\left (  \frac{F_{\rm H{\alpha}}}{2.86F_{\rm H{\beta}}}\right )},
\end{equation}
where $F_{\rm H \alpha}$ and $F_{\rm H \beta}$ are the fluxes of the respective Balmer lines, and the coefficients assume a \cite{Cardelli1989} extinction curve and an unextinguished flux ratio of 2.86 for case B recombination. Then, the corresponding SFR (in M$_{\odot}$ yr$^{-1}$) is obtained using \citep{RosaGonzalez2002}
\begin{equation}
{\rm SFR} = 1.6 \times 7.9 \times 10^{-42} F_{\rm H{\alpha}} 10^{\frac{A_{\rm H{\alpha}}}{2.5}},
\label{eq:sfr}
\end{equation}

\noindent which includes a correction factor of 1.6 to move from a Salpeter IMF (as adopted by {\tt PIPE3D}) to the more commonly used Kroupa IMF  \citep{Speagle2014}. We do this to compare our results with those for other galaxy surveys. 
\noindent We use this to compute the star formation rate surface density,  $\Sigma_{\rm SFR}$ in M$_\odot$\,yr$^{-1}$\,kpc$^{-2}$, by dividing by the face-on area corresponding to a 1\arcsec\ spaxel, given the angular diameter distance to the galaxy. In order to produce smooth SFR maps, we process the H${\alpha}$ and H${\beta}$ fluxes applying the following recipe:

\begin{enumerate}
  \item We select pixels with non-NaN values for ${F_{{\rm H\alpha}}}$.
 \item We adopt a minimum H$\alpha$-to-H${\beta}$ flux ratio, ${F_{{\rm H\alpha}}}$/${F_{{\rm H\beta}}}$ , of $2.86$. Therefore, if ${F_{{\rm H\alpha}}}$/${F_{{\rm H\beta}}}<2.86$, we impose ${F_{{\rm H\alpha}}}$/${F_{{\rm H\beta}}}=2.86$ (so $A_{\rm H\alpha}=0$).
 \item If ${F_{{\rm H\beta}}}$ is a NaN value for a given pixel, then we take the average value of $A_{\rm H\alpha}$ (for pixels with $A_{\rm H\alpha}>0.0$) of the whole $A_{\rm H\alpha}$ map.
\end{enumerate}

We obtain the stellar mass surface density, $\Sigma_\star$, from the stellar maps derived by {\tt{PIPE3D}}. We correct the maps from the spatial binning effect by applying the dezonification correction provided by {\tt{PIPE3D}} datacubes. This is to weight the $\Sigma_\star$ maps by the relative contribution to flux in the $V$-band for each spaxel to the flux intensity of the bin in which it is aggregated (for more details, see \citealt{Sanchez2016_2}). Finally, we mask the $\Sigma_\star$ maps to avoid the flux contribution from field stars and we include the 1.6 correction factor to move from a Salpeter to a Kroupa IMF.

The molecular gas surface density, $\Sigma_{\rm mol}$, is derived from the integrated CO intensity, $I_{\rm CO(2-1)}$, by adopting a constant CO-to-H$_2$ conversion factor that is based on observations of the Milky Way: $X_{\rm CO} = 2 \times 10^{20}$ cm$^{-2}$ (K km s$^{-1}$)$^{-1}$, or equivalently $\alpha_{\rm CO,MW} = 4.3$ M$_\odot$ $(\rm K \, km \, s^{-1} \, pc^{2})^{-1}$ for the CO($J$=1-0) line (\citealt{Walter2008}). We also test how our results depend on our adopted prescription by using the CO-to-H$_2$ conversion factor from Equation 31 in \cite{Bolatto2013}:
\begin{equation}
\alpha_{\rm CO}(Z', \Sigma_{\rm total})  = 2.9 \exp \left ( \frac{+0.4}{Z' \Sigma^{100}_{\rm GMC}} \right ) \left ( \frac{\Sigma_{\rm total}}{100 \, \rm M_\odot \, pc^{2} } \right )^{-\gamma},
\label{eq:alpha_co}
\end{equation}

\noindent in M$_\odot$ $\rm (K \, km s^{-1} \, pc^{-2} )^{-1}$. $\gamma \approx 0.5$ for $\Sigma_{\rm total} > 100$ M$_\odot$ pc$^{-2}$ and $\gamma =0$ otherwise. Additionally, the metallicity is normalized to the solar one, ${Z}' = \rm [O/H]/[O/H]_\odot$, where $\rm [O/H]_\odot = 4.9\times 10^{-4}$ \citep[][]{Baumgartner&Mushotzky2006}, $\Sigma^{100}_{\rm GMC}$ is the average surface density of molecular gas in units of 100 M$_\odot$ pc$^{-2}$, and $\Sigma_{\rm total}$ is the combined gas plus stellar surface density on kpc scales. We are  also interested in the global variations of $\alpha_{\rm CO}({Z}',\Sigma_{\rm total})$.
\noindent To do so, we adopt $\Sigma^{100}_{\rm GMC}=1$ and derive $Z'$ using the metallicity-stellar mass relation (based on the O3N2 calibrator; \citealt{Marino2013}) for CALIFA galaxies from \cite{Sanchez2017}. We use the following expression to obtain $\Sigma_{\rm mol}$
\begin{equation}
\Sigma_{\rm mol}  = \frac{\alpha_{\rm CO}}{R_{21}} \cos (i)\, I_{\rm CO(2-1)},
\label{eq:mol}
\end{equation}

\noindent which adopts the average line luminosity ratio of $R_{21}=I_{\rm CO(2-1)}/I_{\rm CO(1-0)}=0.65$ based on \cite{Leroy2013} and \cite{denBrok2021}, measured at kpc scales; 
\noindent $i$ is the inclination of the galaxy. This equation takes into account the mass correction due to the cosmic abundance of helium. Although $i$ is generally drawn from HyperLEDA, we use SDSS $z$-band images to obtain a better constraint on the inclination (particularly for galaxies with $i>60^\circ$). To do so, we fit an ellipse to the SDSS $z$-band contour for a major axis $A_{\rm maj}\sim 1.2 r_{25}$. We obtain the ratio between the minor and major axes, $A_{\rm min}/A_{\rm maj}$, and compute the inclination by taking $i=\arccos[A_{\rm min}/A_{\rm maj}]$ (see column 4 in Table \ref{table_1}). This assumes an infinitely thin disk and introduces errors for $i>85^\circ$, but we discard highly inclined galaxies from our analysis anyway since most derived quantities are highly uncertain.

To compute the global values of the molecular gas mass, $M_{\rm mol}$, $M_{\star}$, and SFR ($Q_i$ quantities), we use the following equation:
\begin{equation}
Q_{\rm i}  = \int_{A} \Sigma_{\rm i}(r) dA,
\end{equation}

\noindent where $A$ is the area within a circle defined by the geometrical parameters included in Table \ref{table_1} (with radius $r_{25}$ and centered at the optical center); $\Sigma_{\rm i}$ is the surface densities for the pixels within $A$ and i $=$ SFR, mol, or $\star$.

We integrate the surface densities for the molecular gas, stars (assuming that they are distributed along a thin disk), and the SFR to obtain the stellar mass, molecular gas mass, and the SFR within the bulge:

\begin{equation}
M_{{\rm b},i} = 2 \pi \int^{R_{\rm b}}_{0} \Sigma_{i}(r) \, r dr, 
\label{eq:global}
\end{equation}

\noindent where $R_{\rm b}$ is the bulge radius for the stellar component (see \S \ref{bulge_radii} for more details), and $i=$ mol, $\star$, or SFR. We then calculate the integrated ratios as the ratio of the integrated masses and the SFR.

Finally, we compute the resolved SFE$_{\rm mol}$ (in units of yr$^{-1}$) for each pixel, 

\begin{equation}
{\rm SFE_{\rm mol}} = \frac{\Sigma_{\rm SFR}}{\Sigma_{\rm mol}}.
\label{eq:sfe}
\end{equation}

\noindent In a similar way, we calculate the resolved molecular-to-stellar mass fraction, $rR^{\rm mol}_{\star}=\Sigma_{\rm mol}/\Sigma_{\star}$, and the specific star formation rate, sSFR$=\Sigma_{\rm SFR}/\Sigma_{\star}$ (in units of yr$^{-1}$).


\begin{figure*}
  \includegraphics[width=18.2cm]{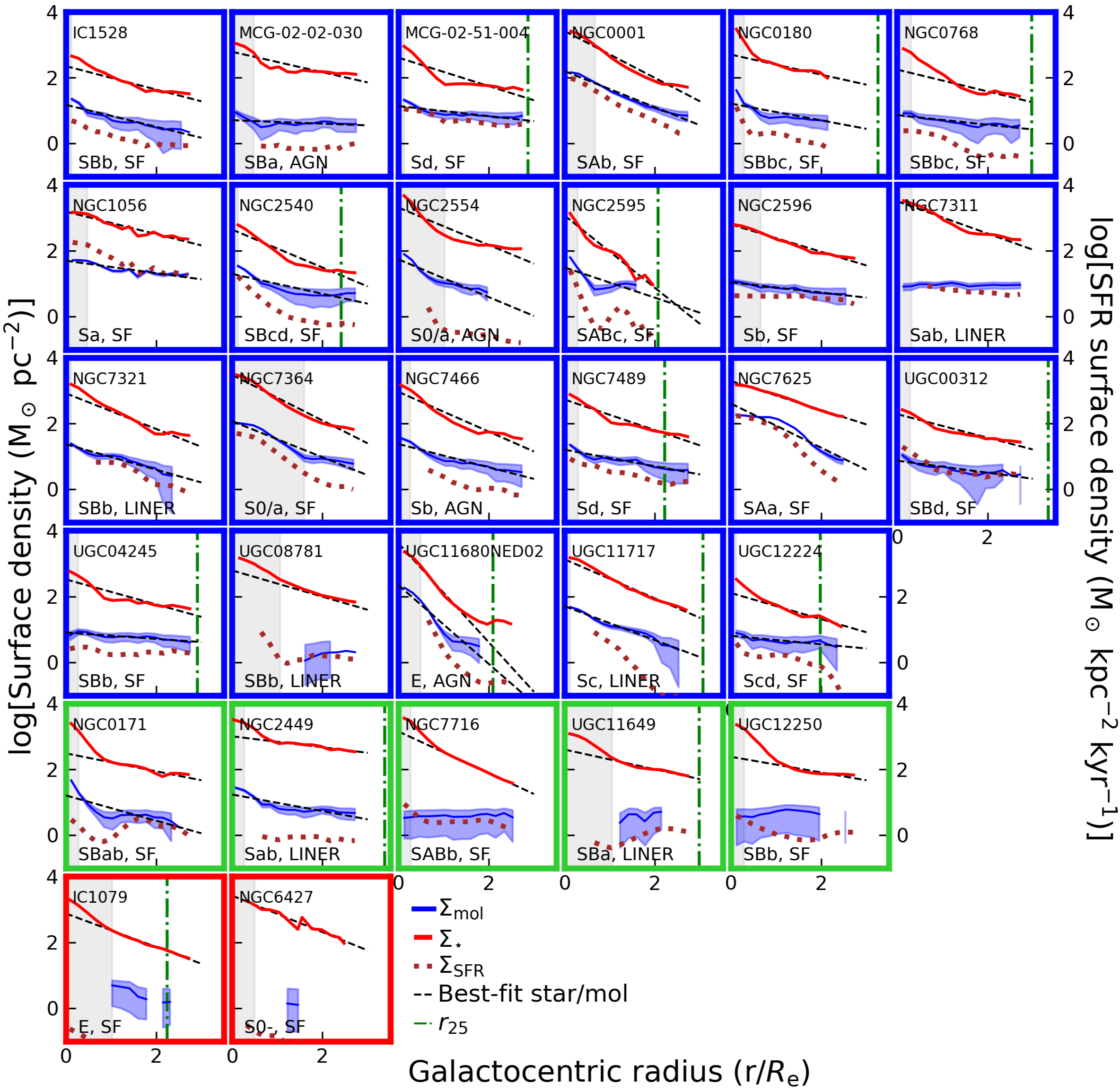}
  \caption{Stellar ($\Sigma_\star$; red solid line) and molecular gas ($\Sigma_{\rm mol}$; blue solid line) surface densities, in units of M$_\odot$ pc$^{-2}$, as a function of galactocentric radius, in units of the stellar effective radius ($R_{\rm e}$), for the 30 ACA EDGE galaxies with $5\sigma$ CO detections and inclinations $i\leq70^\circ$. The blue shaded area is the $\pm 1\sigma$ uncertainty. The brown dotted line is the SFR surface density, $\Sigma_{\rm SFR}$. The gray shaded area is the region within the bulge radius, $R_{\rm bulge}$. Dashed black lines correspond to the best exponential function fit for stellar and molecular gas radial profiles, from top to bottom. The dashed green line corresponds to $r=r_{25}$. The code on the left bottom corner corresponds to the Hubble type and the nuclear activity (columns 5 and 6 in Table \ref{table_2}, respectively). SFRs at $r<0.5 R_{\rm e}$ have been removed for LINER and AGN galaxies since H$\alpha$ in this region is susceptible to LINER/AGN contamination (see \S\ref{SFE_bulge}). Galaxies are classified based on their $\Delta$SFMS as defined in \S \ref{global_relations}, i.e., in main sequence (blue panels), green valley (green panels), and red cloud (red panels). When using stellar profiles as a benchmark, we note a systematic flattening of the molecular gas profiles with $\Delta$SFMS. See also Fig. \ref{fig_8}.}
  \label{fig_profiles}
\end{figure*}

\subsection{Radial profiles}
\label{radial_profiles}

\begin{table*}
\hspace{-2.4cm}
\resizebox{1.13\linewidth}{!}{ 
\begin{tabular}{cccccccccccc}
\hline\hline
Name & log[$M_{\star}$] &  log[SFR]  & log[$M_{\rm mol}$] & Morph. Class & Nuclear & $\Delta$SFR Class & Quenching Stage & $l_\star$ & $l_{\rm CO}$ & $R_{\rm b}$ & $M_{\rm b}$  \\
& (M$_{\odot}$) & (M$_{\odot}$ yr$^{-1}$) & (M$_{\odot}$) & & & & & (kpc) & (kpc) & (kpc) & ($M_\odot /M_\star$)\\
 (1) & (2) & (3) & (4) & (5) & (6) & (7) & (8) & (9) & (10) & (11) & (12) \\
\hline
CGCG429-012 & 10.46 & -1.68 & $<$6.77 &  E0 & ... & RS & fR & ... & ... & ... & ... \\ 
IC1079 & 11.2 & -1.23 & 8.86$\pm$0.17 &  E & ... & RS & nR & 9.12$\pm$0.98 & ... & 10.99$\pm$1.07 & 0.14$\pm$0.19 \\ 
IC1528 & 10.76 & 0.23 & 8.83$\pm$0.02 &  SBb & ... & MS & SF & 4.8$\pm$2.01 & 5.13$\pm$1.25 & 0.52$\pm$0.39 & 0.01$\pm$0.38 \\ 
IC2341 & 10.86 & -0.8 & $<$7.4 &  S0- & ... & RS & fR & ... & ... & ... & ... \\ 
MCG-01-01-012 & 11.19 & 0.21 & 7.58$\pm$0.02 &  SAab & ... & GV & MX & ... & ... & ... & ... \\ 
MCG-01-10-015 & 9.95 & -0.01 & 8.32$\pm$0.03 &  Sd & ... & MS & SF & ... & ... & ... & ... \\ 
MCG-01-52-012 & 10.37 & -0.79 & $<$6.56 &  S0- & ... & GV & nR & ... & ... & ... & ... \\ 
MCG-02-02-030 & 10.81 & 0.03 & 8.52$\pm$0.02 &  SBa & AGN & MS & QnR & 3.95$\pm$1.43 & 23.27$\pm$49.75 & 1.35$\pm$0.28 & 0.1$\pm$0.07 \\ 
MCG-02-51-004 & 10.94 & 0.64 & 9.19$\pm$0.02 &  Sd & ... & MS & SF & 6.93$\pm$0.74 & 19.56$\pm$5.93 & 0.87$\pm$0.69 & 0.02$\pm$0.36 \\ 
NGC0001 & 10.84 & 0.57 & 9.87$\pm$0.02 &  SAb & SF & MS & SF & 2.01$\pm$0.28 & 2.6$\pm$0.47 & 2.25$\pm$0.33 & 0.39$\pm$0.15 \\ 
NGC0155 & 11.08 & -1.51 & $<$7.45 &  S0 & ... & RS & fR & ... & ... & ... & ... \\ 
NGC0169 & 11.24 & 0.73 & 9.54$\pm$0.01 &  SAab & LINER & MS & MX & ... & ... & ... & ... \\ 
NGC0171 & 10.77 & -0.12 & 8.98$\pm$0.02 &  SBab & ... & GV & cQ & 6.94$\pm$4.06 & 4.94$\pm$1.06 & 1.27$\pm$0.44 & 0.15$\pm$0.05 \\ 
NGC0180 & 11.08 & 0.21 & 9.62$\pm$0.02 &  SBbc & ... & MS & QnR & 8.50$\pm$3.90 & 10.16$\pm$2.15 & 1.74$\pm$0.60 & 0.12$\pm$0.05 \\ 
NGC0693 & 9.84 & -0.65 & 6.86$\pm$0.02 &  S0/a & ... & MS & SF & ... & ... & ... & ... \\ 
NGC0731 & 10.94 & -1.55 & $<$6.84 &  E & ... & RS & fR & ... & ... & ... & ... \\ 
NGC0768 & 10.86 & 0.27 & 9.25$\pm$0.02 &  SBbc & ... & MS & SF & 7.92$\pm$2.74 & 17.97$\pm$6.58 & 1.87$\pm$0.60 & 0.13$\pm$0.09 \\ 
NGC0955 & 10.11 & -1.08 & 6.53$\pm$0.04 &  Sab & ... & GV & nR & ... & ... & ... & ... \\ 
NGC1056 & 10.28 & 0.16 & 8.6$\pm$0.01 &  Sa & SF & MS & SF & 1.47$\pm$1.20 & 2.66$\pm$3.95 & 0.56$\pm$0.12 & 0.08$\pm$0.04 \\ 
NGC1542 & 10.57 & -0.03 & 6.34$\pm$0.47 &  Sab & SF & MS & MX & ... & ... & ... & ... \\ 
NGC2449 & 11.13 & -0.2 & 9.03$\pm$0.02 &  Sab & LINER & GV & MX & 6.97$\pm$10.57 & 4.63$\pm$2.16 & 0.72$\pm$0.27 & 0.03$\pm$0.16 \\ 
NGC2540 & 10.54 & 0.11 & 9.47$\pm$0.03 &  SBcd & ... & MS & SF & 4.23$\pm$0.42 & 8.09$\pm$1.54 & 0.74$\pm$0.56 & 0.02$\pm$0.37 \\ 
NGC2554 & 11.11 & 0.68 & 9.4$\pm$0.02 &  S0/a & AGN & MS & nR & 3.89$\pm$0.43 & 3.85$\pm$0.68 & 5.25$\pm$0.51 & 0.47$\pm$0.35 \\ 
NGC2595 & 10.9 & 0.12 & 9.57$\pm$0.02 &  SABc & ... & MS & QnR & 2.96$\pm$0.19 & 7.07$\pm$1.45 & 2.17$\pm$0.75 & 0.24$\pm$0.03 \\ 
NGC2596 & 10.76 & 0.26 & 9.13$\pm$0.02 &  Sb & ... & MS & SF & 4.84$\pm$0.80 & 12.95$\pm$6.32 & 3.07$\pm$0.46 & 0.16$\pm$0.07 \\ 
NGC3300 & 10.76 & -1.75 & $<$6.84 &  SAB & ... & RS & fR & ... & ... & ... & ... \\ 
NGC6427 & 10.63 & -1.85 & 7.39$\pm$0.17 &  S0- & ... & RS & fR & 1.51$\pm$0.57 & ... & 0.97$\pm$0.19 & 0.14$\pm$0.05 \\ 
NGC7025 & 11.17 & 0.23 & $<$7.6 &  Sa & LINER & MS & nR & ... & ... & ... & ... \\ 
NGC7194 & 11.25 & -1.14 & $<$7.86 &  E & LINER & RS & fR & ... & ... & ... & ... \\ 
NGC7311 & 11.12 & 0.36 & 9.1$\pm$0.02 &  Sab & LINER & MS & cQ & 2.77$\pm$0.58 & ... & 0.99$\pm$0.31 & 0.06$\pm$0.14 \\ 
NGC7321 & 11.13 & 0.53 & 9.41$\pm$0.02 &  SBb & LINER & MS & QnR & 4.86$\pm$1.02 & 6.8$\pm$1.15 & 0.72$\pm$0.61 & 0.02$\pm$0.41 \\ 
NGC7364 & 11.18 & 0.66 & 9.58$\pm$0.02 &  S0/a & ... & MS & SF & 2.18$\pm$0.29 & 2.76$\pm$0.47 & 5.55$\pm$0.35 & 0.75$\pm$0.47 \\ 
NGC7466 & 10.95 & 0.36 & 9.56$\pm$0.02 &  Sb & AGN & MS & SF & 4.05$\pm$0.38 & 6.91$\pm$1.10 & 1.51$\pm$0.56 & 0.10$\pm$0.13 \\ 
NGC7489 & 10.83 & 0.54 & 9.42$\pm$0.02 &  Sd & SF & MS & SF & 5.27$\pm$0.64 & 9.78$\pm$2.21 & 1.68$\pm$0.57 & 0.07$\pm$0.11 \\ 
NGC7625 & 10.32 & 0.21 & 9.33$\pm$0.02 &  SAa & SF & MS & SF & 1.13$\pm$0.81 & 0.74$\pm$0.35 & 0.14$\pm$0.11 & 0.01$\pm$0.43 \\ 
NGC7716 & 10.69 & -0.18 & 7.86$\pm$0.08 &  SABb & ... & GV & cQ & 1.69$\pm$0.90 & ... & 0.69$\pm$0.25 & 0.20$\pm$0.10 \\ 
UGC00312 & 10.13 & 0.24 & 8.42$\pm$0.06 &  SBd & SF & MS & SF & 4.45$\pm$1.09 & 8.72$\pm$4.18 & 1.08$\pm$0.37 & 0.04$\pm$0.14 \\ 
UGC00335NED02 & 10.66 & -1.26 & $<$7.58 &  E & LINER & RS & nR & ... & ... & ... & ... \\ 
UGC01123 & 10.73 & -0.87 & $<$6.73 &  Sab & ... & RS & MX & ... & ... & ... & ... \\ 
UGC01368 & 11.21 & 0.53 & 8.93$\pm$0.03 &  Sab & ... & MS & SF & ... & ... & ... & ... \\ 
UGC01938 & 10.64 & 0.13 & 8.68$\pm$0.03 &  Sbc & ... & MS & SF & ... & ... & ... & ... \\ 
UGC02099 & 11.14 & -0.49 & $<$7.58 &  S0 & ... & RS & MX & ... & ... & ... & ... \\ 
UGC04240 & 10.98 & 0.37 & 9.19$\pm$0.02 &  S & ... & MS & SF & ... & ... & ... & ... \\ 
UGC04245 & 10.54 & -0.14 & 9.02$\pm$0.02 &  SBb & ... & MS & cQ & 6.04$\pm$0.95 & 23.75$\pm$14.75 & 1.45$\pm$0.53 & 0.05$\pm$0.14 \\ 
UGC04455 & 11.5 & 0.26 & $<$7.51 &  SBa & ... & GV & cQ & ... & ... & ... & ... \\ 
UGC05396 & 10.81 & 0.23 & 8.36$\pm$0.05 &  Scd & SF & MS & SF & ... & ... & ... & ... \\ 
UGC08322 & 11.15 & 0.06 & 8.67$\pm$0.05 &  Sa & ... & GV & MX & ... & ... & ... & ... \\ 
UGC08781 & 11.09 & 0.37 & 8.65$\pm$0.13 &  SBb & LINER & MS & cQ & 7.29$\pm$1.59 & ... & 7.10$\pm$0.67 & 0.23$\pm$0.12 \\ 
UGC10972 & 10.75 & -0.19 & 8.26$\pm$0.058 & Scd & LINER & GV & cQ & ... & ... & ... & ... \\ 
UGC11649 & 10.57 & -0.47 & 7.76$\pm$0.19 &  SBa & LINER & GV & MX & 6.12$\pm$3.30 & ... & 4.45$\pm$0.42 & 0.08$\pm$0.05 \\ 
UGC11680NED02 & 11.17 & 1.0 & 9.82$\pm$0.02 &  E & AGN & MS & SF & 1.55$\pm$0.17 & 1.95$\pm$0.18 & 2.64$\pm$0.54 & 0.52$\pm$0.12 \\ 
UGC11717 & 11.28 & 0.61 & 9.6$\pm$0.02 &  Sc & LINER & MS & MX & 3.79$\pm$0.73 & 4.36$\pm$0.58 & 0.65$\pm$0.52 & 0.02$\pm$0.41 \\ 
UGC11792 & 10.64 & 0.02 & 8.67$\pm$0.02 &  Sc & SF & MS & SF & ... & ... & ... & ... \\ 
UGC11958 & 11.2 & -0.4 & $<$7.22 &  E & LINER & RS & nR & ... & ... & ... & ... \\ 
UGC11982 & 10.04 & -0.36 & 7.53$\pm$0.12 &  SBcd & ... & MS & SF & ... & ... & ... & ... \\ 
UGC12224 & 10.18 & -0.3 & 9.22$\pm$0.02 &  Scd & SF & MS & SF & 5.94$\pm$1.91 & 18.32$\pm$10.46 & 0.67$\pm$0.54 & 0.02$\pm$0.36 \\ 
UGC12250 & 11.06 & 0.08 & 8.65$\pm$0.1 &  SBb & ... & GV & cQ & 10.39$\pm$5.26 & ... & 1.71$\pm$0.57 & 0.17$\pm$0.09 \\ 
UGC12274 & 11.09 & -0.69 & 8.47$\pm$0.09 &  S & LINER & RS & nR & ... & ... & ... & ... \\ 
UGC12348 & 11.0 & 0.81 & 9.05$\pm$0.02 &  Sa & AGN & MS & MX & ... & ... & ... & ... \\ 
VV488NED02 & 10.9 & 0.36 & 8.43$\pm$0.03 &  SBc & SF & MS & SF & ... & ... & ... & ... \\  
\hline
\end{tabular}}
\caption{\footnotesize Main properties of the ACA EDGE galaxies. Column (1): galaxy name. Column (2) and (3): logarithmic of the total stellar masses and SFRs from CALIFA. Column (4): logarithmic of the total molecular gas mass as derived explained in \S \ref{Basic_equations}. Column (5): morphological classification drawn from NED. Column (6): emission line diagnostics for the optical nucleus spectrum for CALIFA galaxies by \cite{Garcia-Lorenzo2015}, who classify them into star-forming (SF), active galactic nuclei (AGN), and LINER-type galaxies. These groups are also complemented by the type-I and type-II AGN classification by \cite{Lacerda2020}. Column (7): Galaxy classification according to $\Delta$SFR as explained in \S \ref{SFR_Mmol}: main sequence (MS), green valley (GV), and red cloud (RS). Column (8): Two-dimensional emission line classification from \cite{Kalinova2021, Kalinova2022}, who classify galaxies in star-forming (SF), quiescent-nuclear-ring (QnR), centrally quiescent (cQ), mixed (MX), nearly retired (nR), and fully retired (fR). Columns (9) and (10): exponential scale lengths of the molecular gas and stars, respectively, as derived in \S \ref{Exponential_scale}. Columns (11) and (12): radius and mass of bulges, as derived in \S \ref{bulge_radii}.}
\label{table_2}
\end{table*}

We obtain stellar and molecular gas radial profiles for a subsample of 30 galaxies with inclinations $\leq70^{\circ}$ and $5\sigma$ integrated CO detections (see \S \ref{global_relations}), which represent well the distributions of stellar masses and morphologies of the full ACA EDGE sample (see Table \ref{table_2}). We also select spaxels with $3\sigma$ CO detections. We derive these profiles by measuring the average azimuthal CO, stellar, and SFR surface densities in elliptical annuli in the CO(2-1) datacubes. Figure \ref{fig_profiles} shows the molecular gas radial profiles (blue solid line) and their $\pm 1\sigma$ uncertainties (blue shaded areas), which are corrected by inclination (i.e., multiplied by a factor of cos$(i)$). Annuli are centered on the optical galaxy position and aligned with the centered major-axis position angle (column 5 in Table \ref{table_1}). We compute the average $I_{\rm CO(2-1)}$ for a given annulus by summing the velocity-integrated CO line emission intensities from the total pixels within an annulus $\sim$5\arcsec \, wide (average of the minor beam axes), and then we use Equation \ref{eq:mol} (adopting  the constant $\alpha_{\rm CO}$ prescription; see \S \ref{Basic_equations}) to obtain the molecular gas surface density, $\Sigma_{\rm mol}$.

We implement the same method (averaging over all pixels in an annulus) for the SFR and stars. Stellar and SFR radial profiles are shown in Fig. \ref{fig_profiles} \, by the red solid and brown dotted lines, respectively. We remove SFRs at $r<0.5 R_{\rm e}$ for galaxies classified as LINER or AGN (bottom-left corner legend in Fig. \ref{fig_profiles}) since H$\alpha$ in this region is susceptible to LINER/AGN contamination (see \S\ref{SFE_bulge} for more details).

\subsection{Bulge radii and masses}
\label{bulge_radii}

In order to test the star-formation quenching mechanisms within the bulge region (see \S\ref{SFE_bulge}), we derive the radius of the bulge, $R_{\rm b}$, for the 30 ACA EDGE galaxies included in Figure \ref{fig_profiles}. We characterize the bulge-dominated region by identifying the galactocentric radius where there is a break with respect to the stellar radial profiles. Similar to \cite{Villanueva2022}, we adopt $R_{\rm b}=1$ kpc for spiral galaxies where we do not identify a clear break or they have a predominant bar (e.g., SB galaxies have stellar radial profiles mostly dominated by bars and stellar disks rather than bulges). Since previous studies have shown that bulges for spirals are typically $\lesssim 1.5$ kpc (e.g., \citealt{Regan2001,Mendez-Abreu2017,Villanueva2021}), we use the stellar and SFR maps at CALIFA's native resolution of 2.7\arcsec\ to obtain the best physical resolution available ($\sim 0.9$ kpc at the median distance of ACA EDGE galaxies). Bulge radius distributions for main sequence and green valley galaxies are centered at log[$R_{\rm b}/($kpc$)]\sim 0.15$ and $\sim -0.1$, which are slightly larger than those found by \cite{Querejeta2021}, who compute the radius for the central regions (including small bulges and nuclei) of 74 galaxies nearby galaxies selected from PHANGS. By implementing a photometric decomposition using {\tt GALFIT} \citep[][]{Peng2010}, they obtain a mean value of log[$R_{\rm center}/r_{25}]\sim -1.5$, which on average is lower than that of our main sequence galaxies (log[$R_{\rm center}/r_{25}]\sim -1.0$). While ACA EDGE attempts to reflect the broad range of galaxy morphologies in the local universe, PHANGS emphasizes late-type spirals of somewhat lower mass ($9.25\leq $log$[M_\star /$M$_\odot ]\leq 11.25$), which could result in shorter bulge radii. Finally, red cloud galaxies have the largest bulge radii, with $R_{\rm b}$ distributions centered at at log[$R_{\rm B}/($kpc$)]\sim 0.35$.
\noindent These results are consistent with observational evidence. For instance, \cite{Mendel2014} present a catalog of bulge, disk, and total stellar mass estimates for $\sim 660,000$ galaxies from SDSS DR7, based on $g$ and $r$-band photometry published in \cite{Simard2011} and using {\tt GIM2D} \citep[][]{Simard2002}. By fitting S\'ersic profiles \citep[$n_{\rm S}$; ][]{Sersic1968} to elliptical, disk, and bulge$+$disk, they find a S\'ersic index distribution centered at larger values for the former ($n_{\rm S}\sim$5) when compared to the latter two groups ($n_{\rm S}\sim$1). In addition, \cite{Mendez-Abreu2017} implement a 2D photometric decomposition using {\tt GASP2D} \citep[][]{Mendez-Abreu2008,Mendez-Abreu2014} for 404 CALIFA galaxies using $g$, $r$, and $i$ SDSS images, including 28 ACA EDGE galaxies in their analysis. We obtain a close 1:1 relation when comparing the two sets of bulge radii (OLS $R_{\rm b,CALIFA}=[0.83\pm0.10]\times R_{\rm b,ACA EDGE}$), which also show a strong correlation between them (Pearson $r_{\rm p}=0.92$; $p$-value$<<0.01$).

\begin{figure*}
  \includegraphics[width=9.cm]{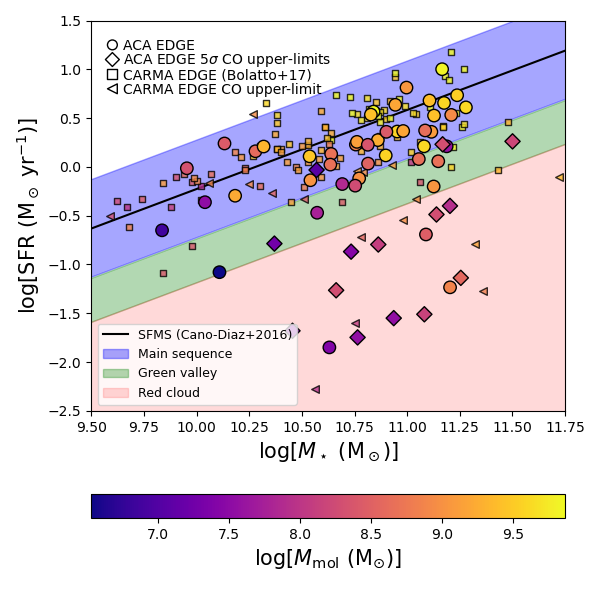} 
  \includegraphics[width=9cm]{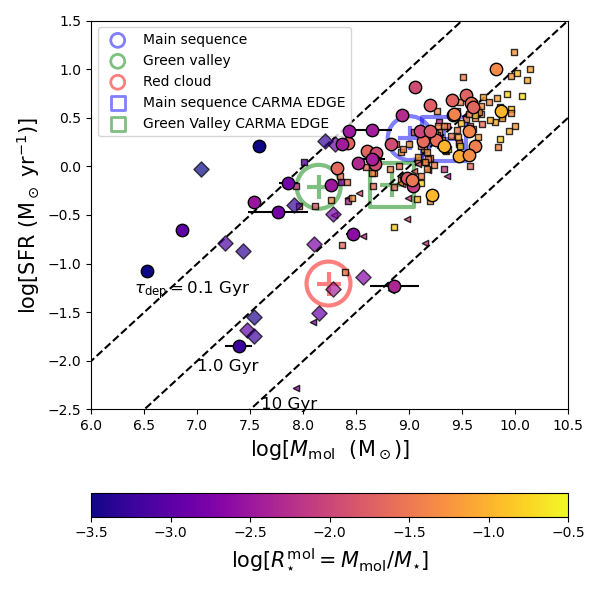} \\
  \includegraphics[width=9.1cm]{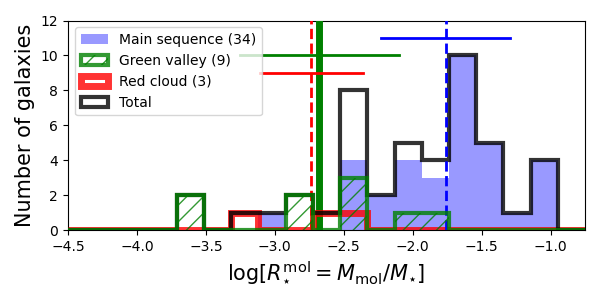} 
  \includegraphics[width=9cm]{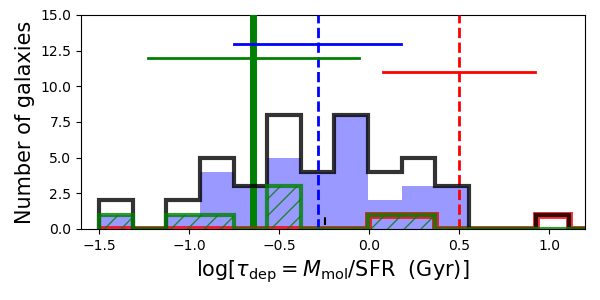} \\
  \caption{{\it Top left:} SFR-$M_{\star}$ diagram integrated over CALIFA SFR and stellar maps, color-coded by the total molecular gas mass, $M_{\rm mol}$. The solid black line is the SFMS fit by \cite{Cano-Diaz2016}. Blue, green, and red dashed areas define main sequence, green valley, and red cloud galaxies, respectively, as defined by the bands (see \S \ref{global_relations} for more details). {\it Top right:} SFR-$M_{\rm mol}$ relation color-coded by stellar mass. The dashed black lines, from top to bottom, correspond to molecular gas depletion times $\tau_{\rm dep} = 0.1$, $1.0$, and $10$ Gyr, respectively. Blue, green, and red circles are the centroids of SFR and $M_{\rm mol}$ values for galaxies with 5$\sigma$ CO detections (filled circles) of the groups as defined by the bands in the top left panel. The blue and green squares correspond to the centroid of SFR and $M_{\rm mol}$ values for main sequence and green valley CARMA EDGE detected galaxies. {\it Bottom:} Distributions for the molecular-to-stellar mass fraction ($R^{\rm mol}_{\star}=M_{\rm mol}/M_{\star}$; left) and the molecular gas depletion time ($\tau_{\rm mol}=M_{\rm mol}/$SFR; right) for the three categories (excluding CO upper-limits), as defined in top left panel. Vertical and horizontal lines correspond to the average values and the standard deviations of the distributions, respectively. The plots suggest that while the transition from main sequence to green valley galaxies is mostly driven by gas removal/depletion, the movement from the latter to the red cloud may be determined by a reduction in the star formation efficiency of the molecular gas (SFE$_{\rm mol}=\tau^{-1}_{\rm dep}$).}
  \label{fig_3}
\end{figure*}

Using $R_{\rm b}$, we compute the bulge mass, $M_{\rm b}$, in terms of the total stellar mass, after numerically integrating the stellar profiles using Equation \ref{eq:global}. Table \ref{table_2} summarizes the properties of the 60 ACA EDGE galaxies, together with the values of $R_{\rm b}$ and $M_{\rm b}$ (columns 10 and 11). Columns (4), (8), and (9) list $M_{\rm mol}$, $l_{\rm \star}$, and $l_{\rm mol}$, respectively; the latter two are calculated from radial profiles in \S \ref{Exponential_scale}.


\section{Results and Discussion}
\label{S4_Results}

In the next subsections we present the main properties of the 60 galaxies included in the ACA EDGE survey. To do so, we divide our results in global (i.e., quantities derived from integrated measurements) and spatially resolved (i.e., quantities derived from pixel measurements). Unless otherwise mentioned, we estimate the molecular gas related quantities by adopting a constant Milky Way CO-to-H$_2$ conversion factor (see \S \ref{Basic_equations}).

\subsection{Global quantities and relations}
\label{global_relations}

\subsubsection{SFR versus stellar and molecular gas masses}
\label{SFR_Mmol}

The top left panel of Figure \ref{fig_3} shows the SFR-$M_{\star}$ relation, color-coded by $M_{\rm mol}$, using the global values (pixels at $r<r_{25}$) of SFR and $M_{\star}$ (see \S \ref{Basic_equations}). On average, we note that galaxies near the SFMS (black line; \citealt{Cano-Diaz2016}) tend to have higher molecular gas masses, although there is not a clear region on the SFR-$M_{\star}$ relation associated with low values of $M_{\rm mol}$ (see color-coded symbols). In order to characterize the behaviour of the molecular gas as a function of the difference between the SFR and the SFMS, $\Delta$SFMS$=$ log[SFR]$-$SFMS, we classify our galaxies in three different groups based on their $\Delta$SFMS, as shown by shaded areas in top left panel of Figure \ref{fig_3}:

\begin{enumerate}
    
    \item Main sequence (36 galaxies, 34 with $5\sigma$ CO detections; blue shaded area), which are galaxies with $-0.5$ dex$< \Delta$SFMS.
    
    \item Green valley (11 galaxies, 9 with $5\sigma$ CO detections; green shaded area), encompassing galaxies with $-1.0$ dex$< \Delta$SFMS $\leq-0.5$ dex.

    \item Red cloud (13 galaxies, 3 with $5\sigma$ CO detections; red shaded area), which includes galaxies with $\Delta$SFMS $\leq-1.0$ dex.
    
\end{enumerate}

\noindent We choose these boundaries based on the typical values of the main-sequence/green valley distribution scatters reported in recent studies, which span from $\sim0.2$ to $0.7$ dex \citep[e.g.,][]{Schawinski2014,Chang2015,Bland-Hawthorn2016,Cano-Diaz2016,Sanchez2018,Colombo2020}. The bottom left panel of Figure \ref{fig_3} shows the distribution of the molecular-to-stellar mass fraction, $R^{\rm mol}_{\star}=M_{\rm mol}/M_{\star}$, of the three groups for galaxies with $5\sigma$ CO detections. Main sequence galaxies have the highest molecular gas masses (with an average log$[R^{\rm mol}_{\star}]\sim -1.6$ dex; blue dashed line), while on average both green valley and red cloud galaxies have fractions about an order of magnitude lower (green solid and red dashed lines). 

The top right panel of Fig. \ref{fig_3} shows the SFR-$M_{\rm mol}$ relation, color-coded by $M_\star$. The three dashed black lines correspond to three different molecular depletion times, $\tau_{\rm dep}=M_{\rm mol}/$SFR = 0.1, 1.0, and 10 Gyr, going from top to bottom, respectively. It is interesting to note that although most ACA EDGE galaxies are well represented by the $\tau_{\rm dep}= 1$ Gyr line, there is not a characteristic molecular depletion time for the whole sample. This is confirmed when we analyze the molecular gas depletion time distributions of the three groups (bottom right panel of Fig. \ref{fig_3}); red cloud galaxies have $\tau_{\rm dep}$ around 3 and 6 times longer than main sequence and green valley galaxies, respectively. However, these results have to be considered carefully due to the small number of CO-detected red cloud galaxies.

Our results are consistent with \cite{Colombo2020}, who analyze $^{12}$CO($J$=2-1) APEX data at 26.3\arcsec \, resolution (i.e., the region within $R_{\rm e}$) for 472 EDGE galaxies. They note a strong correlation between $\Delta$SFMS and the star formation efficiency of the molecular gas, SFE$_{\rm mol}= \tau^{-1}_{\rm dep}$, suggesting a scenario where the transition of galaxies from the main sequence to the green valley is primarily driven by the molecular gas lost. In addition, they propose that changes in the SFE$_{\rm mol}$ of the remaining cold gas is what modulates a galaxy's retirement from the green valley to the red cloud. Analyzing a compilation of $\sim$8000 galaxies from MaNGA, \cite{Sanchez2018} also note that the SFE decreases as galaxies move out of the main sequence to the red cloud and pass through the green valley, which is confirmed by several studies (e.g.,  \citealt{Sanchez_1_2020,Brownson2020,Sanchez2021RMx,Lin2022}).


\subsubsection{Exponential scale lengths}
\label{Exponential_scale}

\begin{figure}
    \hspace{-0.5cm}
  \includegraphics[width=8.8cm]{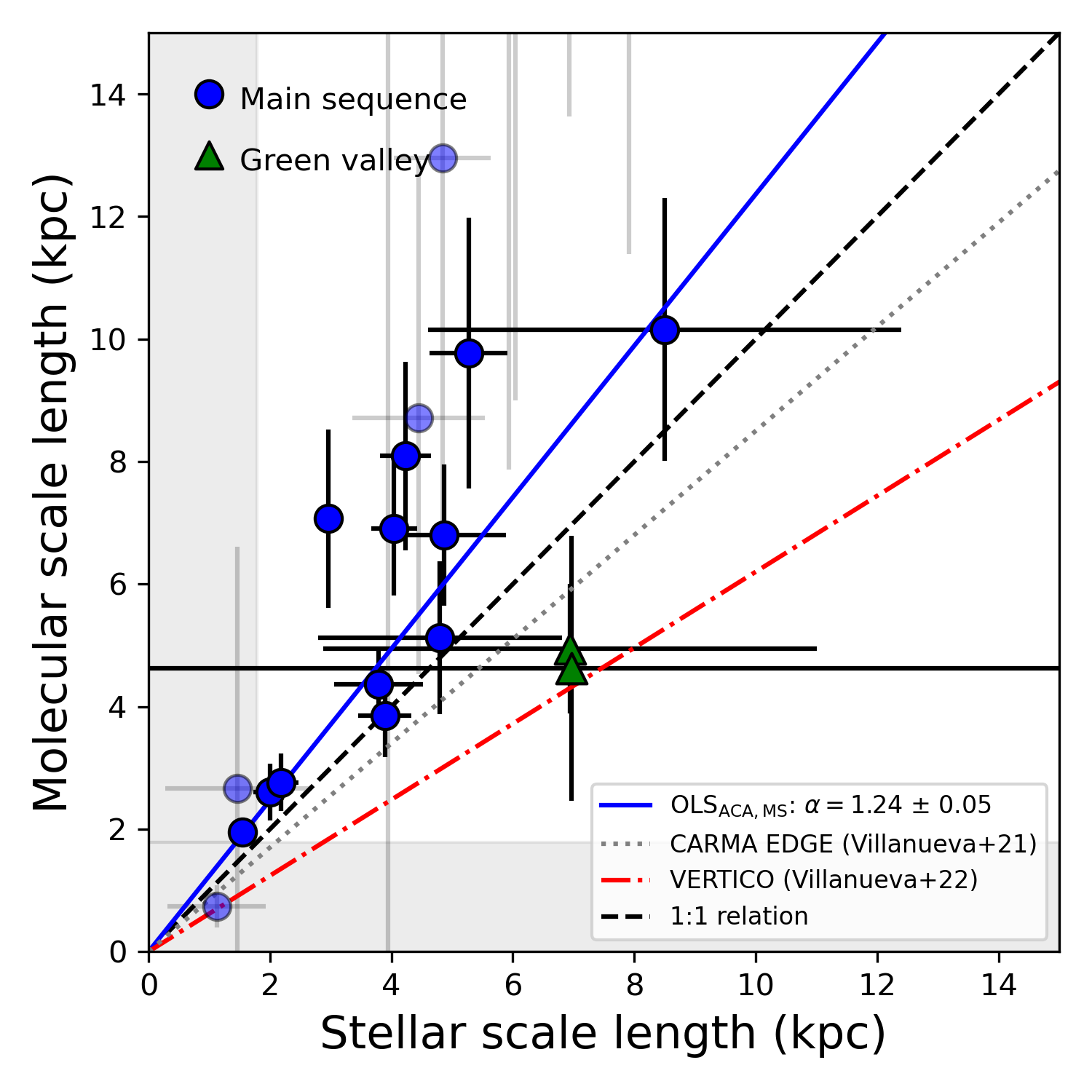} 
  \caption{Comparison between the stellar, $l_{\star}$, and molecular, $l_{\rm mol}$, scale lengths, computed by fitting exponential profiles to the respective surface densities as a function of galactocentric radius. Blue circles and green triangles correspond to 23 ACA EDGE galaxies with $\Sigma_{\rm mol}>1$ M$_{\odot}$ pc$^{-2}$ for all the annuli within 1 kpc. The blue solid line is the best fit for the model $y=\alpha x$ for main sequence, omitting galaxies with low-quality $l_{\rm mol}$ fits (symbols with pale colors). The gray dotted and orange-dotted lines are the best fit relation for CARMA EDGE \citep[][]{Villanueva2021} and VERTICO \citep[][]{Villanueva2022}, respectively. The shaded gray area correspond to the median physical resolution of ACA EDGE galaxies. On average, the figure shows a $\sim$6:5 relation between the molecular and stellar scale lengths.}
  \label{fig_4}
\end{figure}

{If gas removal/depletion is one of the main processes modulating the transition from main sequence to green valley galaxies, it should impact the distribution of the molecular gas. To test this, we compute the exponential scale lengths for the molecular gas, $l_{\rm mol}$, and the stars, $l_{\star}$, for the ACA EDGE galaxies in Figure \ref{fig_profiles}. Figure \ref{fig_4} shows the comparison between $l_{\rm mol}$ and $l_{\star}$ for main sequence ACA EDGE galaxies with $i<70^\circ$ and $5\sigma$ CO detections (see \S \ref{radial_profiles})}. Galaxies are color-coded by their $\Delta$SFMS according to the classification explained in Figure \ref{fig_3}. Out of the 30 galaxies with molecular gas and stellar radial profiles, we have selected 23 galaxies with decreasing $\Sigma_{\rm mol}$ profiles (i.e., $\Sigma_{\rm mol}$($r<$1 kpc)$>\Sigma_{\rm mol}(r=r_{\rm max})$,
\noindent where $r_{\rm max}$ is the largest radius at which we have a $5\sigma$ CO detection). Since we also restrict the $\Sigma_{\star}(r)$ exponential fit to the stellar disk, we do not consider annuli within prominent bulges (i.e., E and S0 galaxies; \citealt{Regan2001}) and bars (i.e., SB, Sab, and Sbc galaxies). These fits for $\Sigma_{\rm mol}$ and $\Sigma_{\star}$ profiles are shown by the black dashed lines in Figure \ref{fig_profiles}. We observe a significant correlation between $l_{\rm mol}$ and $l_{\star}$ for main sequence and green valley galaxies (blue and green circles; Pearson $r_{\rm p}=0.6$; $p$-value $<0.01$). When we compute an ordinary least-square (OLS; blue solid line in Fig. \ref{fig_4}) bisector fit for the model $y=\alpha x$ for main sequence galaxies with at least $5\sigma$ $l_{\rm mol}$ measurements (symbols with enhanced color in Fig. \ref{fig_4}), we obtain $l_{\rm mol}=(1.24\pm 0.05)\times l_{\star}$. We test how this relation varies with angular resolution by fitting the CO radial profiles derived from CO moment 0 maps smoothed at 10\arcsec\ beamsize. Although molecular length scales are slightly larger than for stars $l_{\rm mol}=(1.15\pm 0.05)\times l_{\star}$, the best linear relation is still above unity.

While several studies have found a close 1:1 relation between the molecular gas and stars in main sequence star-forming galaxy samples based on galaxies selected from the field (e.g., \citealt{Young1995}; BIMA \citealt{Regan2001}; HERACLES, \citealt{Leroy2008}; CARMA EDGE, \citealt{Bolatto2017,Villanueva2021}), quenching mechanisms have the potential to affect the distribution of the molecular gas, atomic gas, and stars in different ways. On the one hand, environmental mechanisms (e.g., ram pressure stripping, galaxy interactions, among others) have been shown to compact the spatial extent of the molecular gas, particularly in high-density environments (e.g., galaxy clusters; \citealt{Boselli2014b,Zabel2022}). For instance, \cite{Villanueva2022} find a $\sim$3:5 relation for the molecular and stellar scale lengths in a subsample of 28 Virgo Cluster galaxies selected from VERTICO \citep[][]{Brown2021}. On the other hand, intrinsic mechanisms tend to operate either by removing (e.g., via AGN activity), re-distributing (e.g., via stellar feedback), or depleting (e.g., via starvation) the cold gas reservoirs. Figure \ref{fig_profiles} shows a broad variety of radial profiles that could be explained by a different combination of mechanisms depending on the galaxy $\Delta$SFMS. The best relation between molecular gas and stellar scale lengths for ACA EDGE main sequence galaxies (blue solid circles in Fig. \ref{fig_4}) {is close to a 6:5 relation. Although this is still consistent with the almost $\sim$1:1 relation from \cite{Villanueva2021}, $l_{\rm mol}$ values for ACA EDGE galaxies are slightly larger when compared to CARMA EDGE spirals. This seems to be the result of the lower molecular gas content in the central regions of the former rather than the latter (as shown by the $M_{\rm mol}$ centroids in upper-right panel of Figure \ref{fig_3}). This in consequence produces flatter $\Sigma_{\rm mol}$ profiles in ACA EDGE galaxies than those for CARMA EDGE, which were mainly selected to be bright in far-IR (i.e. rich in molecular gas; see \S \ref{sample_selection} and \citealt{Bolatto2017} for more details).}

\begin{figure*}
  \includegraphics[width=18cm]{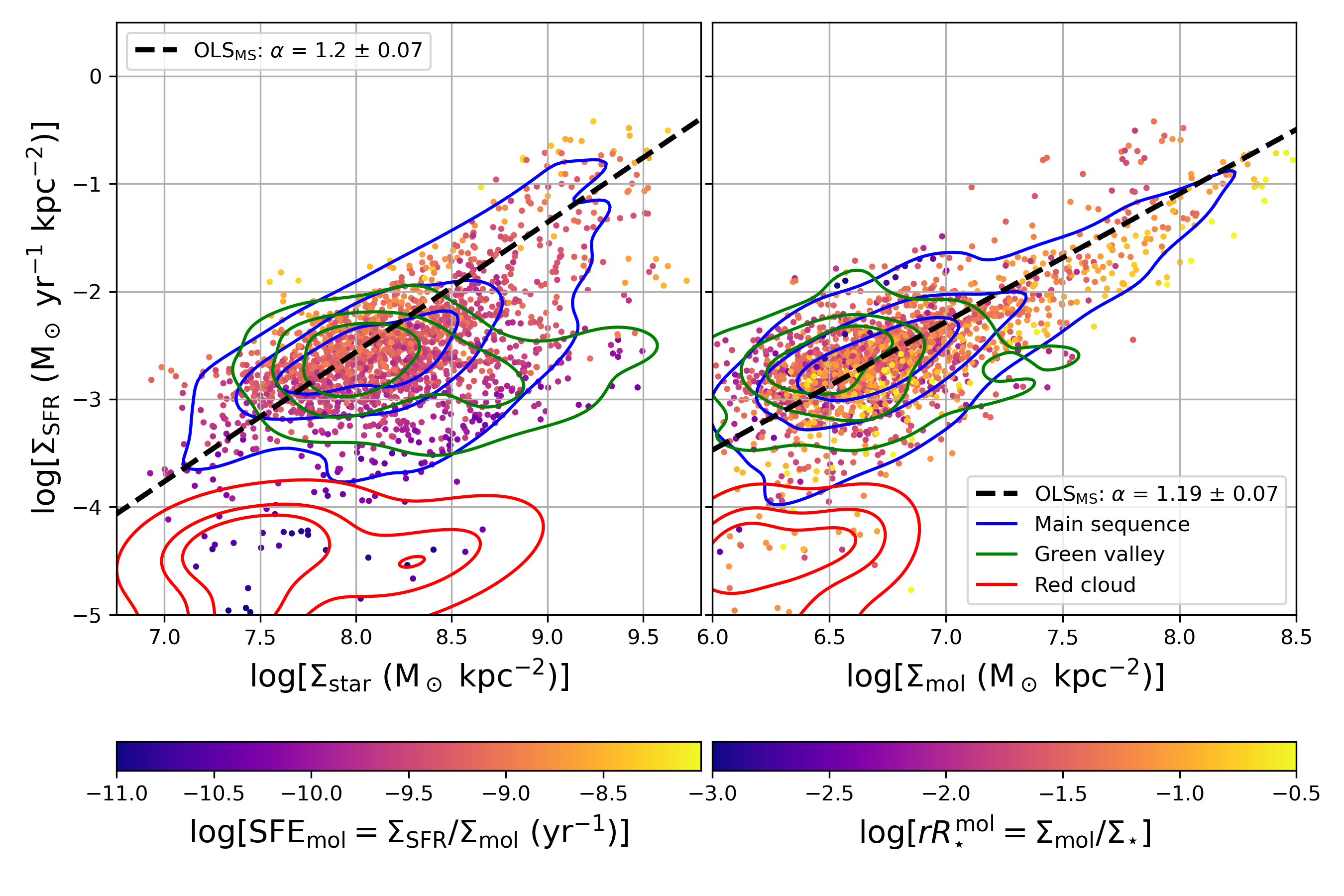} 
  \caption{{\it Left:} SFR surface density, $\Sigma_{\rm SFR}$, versus stellar surface density, $\Sigma_{\star}$, color-coded by the resolved star formation efficiency of the molecular gas, SFE$=\Sigma_{\rm SFR}/\Sigma_{\rm mol}$, for pixels with $5\sigma$ CO detections and selected from the 30 galaxies included in Figure \ref{fig_profiles}. Blue and green density contours are 90\%, 60\%, and 30\% of the points for main sequence and green valley galaxies. 
  \noindent {\it Right:} The resolved SFR-$M_{\rm mol}$ relation, color-coded by the resolved molecular-to-stellar mass gas fraction, $rR^{\rm mol}_{\star}=\Sigma_{\rm mol}/\Sigma_{\star}$. Conventions are as in left panel. The black dashed lines corresponds to the OLS bisector fit for main sequence galaxies using the model $y = \alpha x+ \beta$ for the resolved SFMS (left) and the resolved Kennicutt-Schmidt (right) relations. While the left panel exhibits an increasing in $\Sigma_\star$ for pixels transiting from the main sequence to the green valley, the right panel shows that pixels from these two populations cover a similar parameter space although with a mild decreasing in $\Sigma_{\rm SFR}$. This suggests that changes in star formation activity during the transition are driven not only by a lowering in the molecular gas, but also due to a decrease of the star formation efficiency.}
  \label{fig_6}
\end{figure*}



\subsection{Spatially resolved relations}
\label{resolved_relations}

\subsubsection{SFR versus stellar and molecular gas surface densities}
\label{SFR_Mmol_resolve}

The left panel of Figure \ref{fig_6} shows $\Sigma_{\rm SFR}$ versus $\Sigma_{\star}$ (the so-called resolved SFMS, rSFMS; e.g., \citealt{Cano-Diaz2016,Lin2019,Sanchez2021,Ellison2021b}), both in units of M$_\odot$ kpc$^{-2}$ and color-coded by the resolved star formation efficiency of the molecular gas, SFE$_{\rm mol}=\Sigma_{\rm SFR}/\Sigma_{\rm mol}$. The figure includes pixels from the 30 ACA EDGE galaxies with $5\sigma$ global CO detections and $i<70^\circ$. Similarly to \S \ref{global_relations}, we classify pixels according to the $\Delta$SFMS of the host galaxy as main sequence (blue contours), green valley (green contours), and red cloud (red contours). Although there is not a remarkable difference in the $\Sigma_{\star}$ range covered by the main sequence and green valley pixels, there is a mild decrease in $\Sigma_{\rm SFR}$ from the former (log$[\Sigma_{\rm SFR}]\sim -2.7$ dex) to the latter (log$[\Sigma_{\rm SFR}]\sim -3.0$ dex). However, red cloud pixels have the lowest SFR of all groups. To compare our results with previous studies, we compute an OLS bisector fit for main sequence pixels using the model $y = \alpha x+ \beta$; we obtain $\rm \log[\Sigma_{\rm SFR}]= (1.20\pm0.07)\times \log[\Sigma_{\star}] - (12.18\pm0.60)$ (dashed black line in left panel of Fig. \ref{fig_6}). Our rSFMS best-fit slope, $\alpha_{\rm rSFMS}$, is slightly higher than those for CARMA EDGE ($\alpha_{\rm rSFMS}\approx1.01$; \citealt{Bolatto2017}), PHANGS ($\alpha_{\rm rSFMS}\approx1.04$; \citealt{Pessa2021}), and other galaxy sample (see \citealt{Sanchez2023IAU} and references therein). However, our results are consistent with the values found in several studies based on galaxy samples similar to ACA EDGE (e.g. \citealt{Lin2019}). For instance, \cite{Ellison2021b} analyze the rSFMS properties of $\sim$15,000 spaxels in a sample of 29 galaxies selected from the ALMA-MaNGA QUEnching and STar formation (ALMaQUEST) survey \citep[][]{Lin2020}. Covering the same range of stellar masses, ALMaQUEST was designed to investigate the star-formation activity in galaxies from the green valley to the starburst regime, complementing surveys with a better representation of galaxy properties in the local Universe (e.g., CARMA EDGE). Implementing an orthogonal distance regression (ODR) fit for the rSFMS, \cite{Ellison2021b} find $\rm \log[\Sigma_{\rm SFR}]= (1.37\pm0.01)\times \log[\Sigma_{\star}] - (13.12\pm0.10)$, thus resulting in a steeper rSFMS slope (clearly above unity) for high stellar mass galaxies. 

Similarly to the rSFMS, the widely-studied resolved Kennicutt-Schmidt relation (rKS; e.g., \citealt{Bigiel2008,Leroy2008,Schruba2011,Pessa2021,Sanchez2021b,JimenezDonaire2023,Sun2023}) presents a complementary way to investigate how the SFR depends on the ISM. The right panel of Figure \ref{fig_6} contains the rKS relation for ACA EDGE galaxies, color-coded by the resolved molecular-to-stellar mass fraction, $rR^{\rm mol}_{\star}=\Sigma_{\rm mol}/\Sigma_{\star}$, and density contours as in the left panel. It is interesting to note that the OLS bisector fit for main sequence galaxies also yields a rKS best-fit slope value, $\alpha_{\rm rKS}$, above unity ($\rm \log[\Sigma_{\rm SFR}]= (1.19\pm0.07)\times \log[\Sigma_{\rm mol}] - (10.62\pm0.98)$; dashed black line in right panel of Fig. \ref{fig_6}). Although our $\alpha_{\rm rKS}$ is higher when compared to that for CARMA EDGE ($\alpha_{\rm rKS}\approx1.01$; \citealt{Bolatto2017}), PHANGS ($\alpha_{\rm rKS}\approx1.03$; \citealt{Pessa2021}), and other galaxy samples from the literature (see \citealt{Sanchez2023IAU} and references therein), it is consistent with the ODR fit for ALMaQUEST galaxies ($\rm \log[\Sigma_{\rm SFR}]= (1.23\pm0.01)\times \log[\Sigma_{\rm mol}] - (10.49\pm0.06)$; \citealt{Ellison2021b}). We note however that these results are very sensitive to the adopted $\alpha_{\rm CO}$ prescription. For instance, \cite{Sun2023} show that different assumptions of the CO-to-H$_2$ conversion factor can result in $\alpha_{\rm rKS}=0.9-1.2$, which translates into uncertainties up to 25\% in the CO related quantities of PHANGS galaxies. We also observe a systematic decrease in both $\Sigma_{\rm SFR}$ and $\Sigma_{\rm mol}$ from the main sequence to the green valley galaxies. In combination with the results shown in the left panel, this may suggest that although the transition from main sequence to the green valley is primarily driven by gas removal, a decrease in SFE$_{\rm mol}$ also plays a role in modifying the ability of the molecular gas to form stars (see the color-coded points in left panel of Fig. \ref{fig_7B}).

\subsubsection{SFE and bulge properties}
\label{SFE_bulge}


\begin{figure}
  \includegraphics[width=8.5cm]{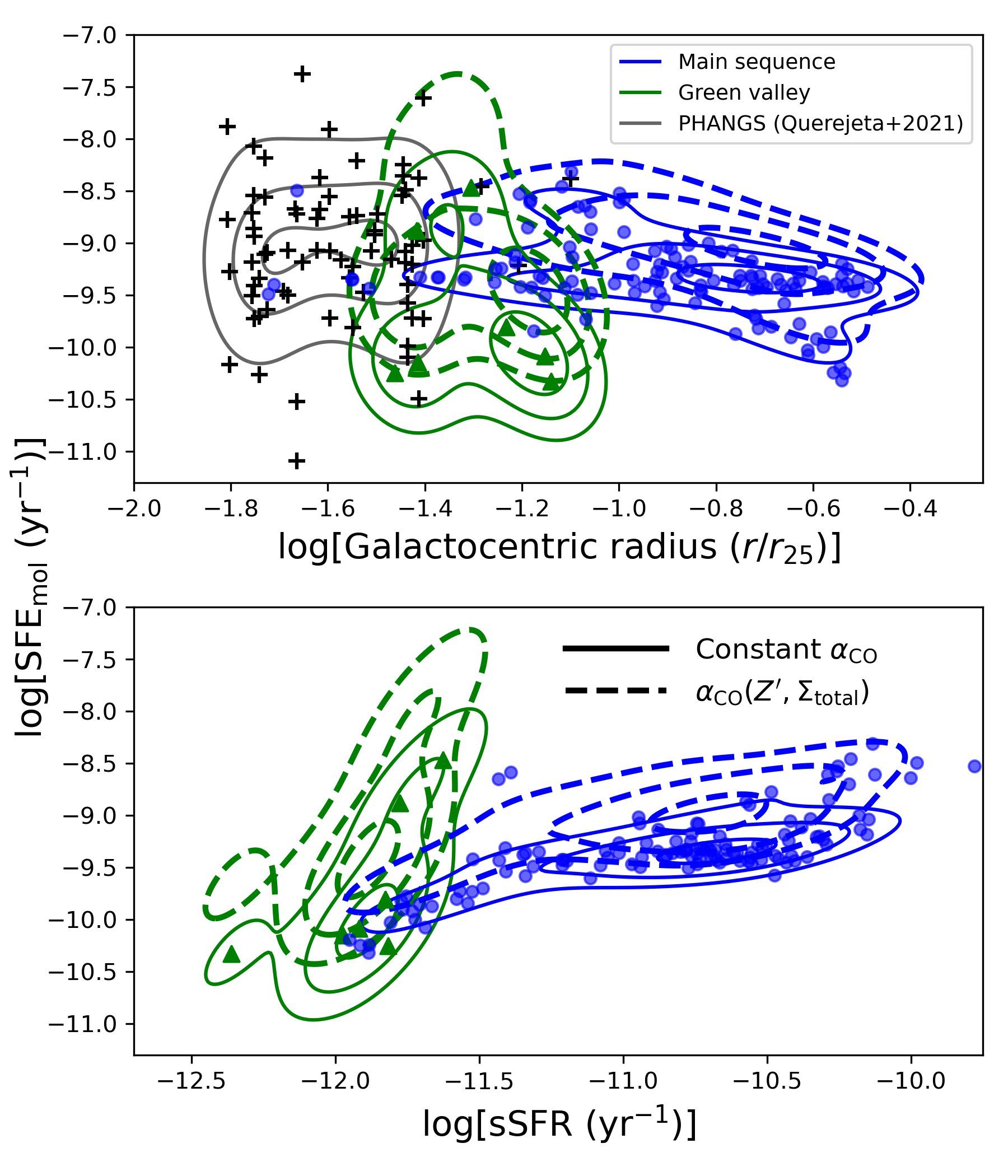} 
  \hspace{0.8cm}
  \includegraphics[width=8.4cm]{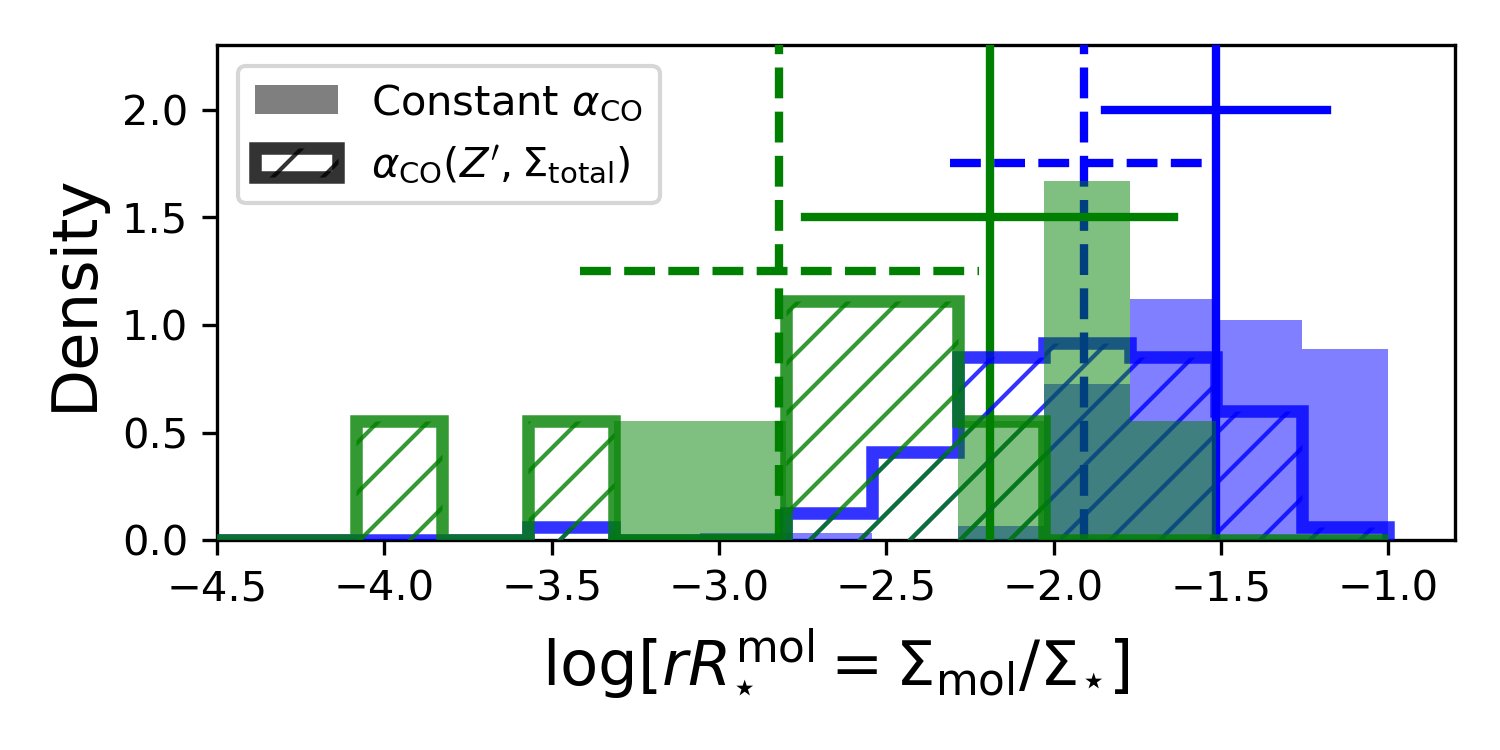} 

  \vspace{-0.4cm}
  \caption{{\it Top:} The resolved SFE$_{\rm mol}$ versus galactocentric radius for SF pixels within $R_{\rm b}$, color-coded by their $\Delta$SFMS. Black crosses are pixels drawn for PHANGS-ALMA spirals. Solid blue and green lines correspond to 90\%, 60\%, and 30\% density contours of main sequence and green valley pixels, respectively. Dashed lines are density contours for pixels when adopting a variable $\alpha_{\rm CO}(Z',\Sigma_{\rm total})$ prescription (see Eq. \ref{eq:alpha_co}). {\it Middle:} The resolved SFE$_{\rm mol}$ versus the resolved specific star formation rate, sSFR, for the same groups as in top panel. {\it Bottom:} Distribution of the resolved molecular-to-stellar mass fraction, $rR^{\rm mol}_{\star}$, for main sequence and green valley galaxies included in the upper panel. The vertical and horizontal lines are the mean and the standard deviation values of the distributions, respectively. We note that the spatially resolved SFE$_{\rm mol}$, sSFR, and $rR^{\rm mol}_{\star}$ within the bulges have a systematic decrease with $\Delta$SFMS, and these trends seem to not depend on the adopted $\alpha_{\rm CO}$ prescription.}
  \label{fig_7B}
\end{figure}

To understand which mechanisms may be driving the star-formation quenching in ACA EDGE galaxies, we analyze the impact of bulges on the star formation efficiency of the molecular gas. 
\noindent It is important to mention that SFR estimators derived from H$\alpha$ have to be taken carefully since they are susceptible to contamination due to AGN activity, jets, shocks and post-Asymptotic Giant Branch stars \citep[][]{Lacerda2020}. To perform our analysis only on star-forming pixels, we have used estimates of the nuclear activity of CALIFA galaxies from \cite{Garcia-Lorenzo2015} (column 5 in Table \ref{table_2}), who classify galaxies according to the emission-line diagnostic of the optical nucleus in star-forming (SF), AGN, and LINER-type galaxies. Although recent studies have proposed the term LIERs (or ``low ionization emission regions'') to redefine the term ``LINER'' since the latter is not only limited to nuclear regions neither restricted to galaxy centers (e.g., \citealt{Singh2013,Belfiore2016}), for simplicity we use hereafter the term LINER. We complement the AGN classification using \cite{Lacerda2020}, who group CALIFA galaxies as Type-I (galaxies with a broad H$\alpha$ width, i.e. FWHM$>1000$ km s$^{-1}$) or Type-II (galaxies above the \citealt{Kewley2001} line on the BPT diagram and H$\alpha$ line width$>3 \rm \AA$) AGNs. Although galaxies may host an AGN and actively form stars, we classify galaxies as SF if no nuclear activity is detected. We adopt this since we do not see significant variations between the results for confirmed SF-only galaxies and SF+not-detected nuclear activity galaxies. 

{The top panel of Figure \ref{fig_7B} shows the resolved SFE$_{\rm mol}$ as a function of galactocentric radius (in units of $r_{25}$), color-coded by the $\Delta$SFMS of the host galaxy, for SF-pixels within $R_{\rm b}$. The figure also includes the SFE$_{\rm mol}$ pixels within the centers (including bulge and nucleus) of PHANGS-ALMA galaxies drawn from \cite{Querejeta2021} (black crosses), which complement the ACA EDGE sample by providing data at smaller galactocentric radii. On average, ACA EDGE green valley pixels have lower efficiencies compared to those for PHANGS and ACA EDGE main sequence galaxies, with the two latter covering a similar range of SFE$_{\rm mol}$. To test how these results depend on the $\alpha_{\rm CO}$ prescription, we compute the SFE$_{\rm mol}$ by adopting a variable $\alpha_{\rm CO}(Z',\Sigma_{\rm total})$ (see Equation \ref{eq:alpha_co}), as shown in the top panel of Figure \ref{fig_7B} by dashed contours. On average, $\alpha_{\rm CO}(Z',\Sigma_{\rm total})$ values are lower than for the fixed prescription at $r\lesssim1.5 R_{\rm e}$ 
\noindent; consequently, SFE$_{\rm mol}$ are higher when derived from $\alpha_{\rm CO}(Z',\Sigma_{\rm total})$. We note that green valley galaxies have a slightly higher increase in the efficiencies than main sequence galaxies ($\sim0.3$ dex) with the variable $\alpha_{\rm CO}$, although with the former still having lower SFE$_{\rm mol}$ than the latter. The middle panel of Figure \ref{fig_7B} shows the SFE$_{\rm mol}$ as a function of the specific star-formation rate, sSFR$=\Sigma_{\rm SFR}/\Sigma_{\star}$, for ACA EDGE pixels within the bulge region. We note a systematic increase of the efficiencies with sSFR, going from low SFE$_{\rm mol}$ values for green valley galaxies (log[SFE$_{\rm mol}$]$\sim$-10.3 and log[sSFR]$\sim$-12), to high SFE$_{\rm mol}$ values for main sequence galaxies (log[SFE$_{\rm mol}$]$\sim$-9.3 and log[sSFR]$\sim$-10.5). Even though efficiencies are higher when compared to those derived from the fixed $\alpha_{\rm CO}$, these tendencies do not change when adopting the variable $\alpha_{\rm CO}$ prescription (as shown by dashed contours in top and middle panels of Figure \ref{fig_7B}). These results are in agreement with several studies reporting lower star formation efficiencies in bulge dominated galaxies (e.g., \citealt{Colombo2018,Ellison2021,Sanchez2021RMx}). For instance, \cite{CatalanTorrecilla2017} report a decrease in the SFRs with sSFR within bulges of CALIFA galaxies at any $M_{\star}$. \cite{Eales2020} also find a clear correlation between the star formation efficiency and sSFR in galaxies without prominent bulges and with the same morphological type. In addition, they note a strong connection between massive bulges and low SFE. 

Are the differences in SFE$_{\rm mol}$ between main sequence and green valley bulges primarily driven by gas depletion/removal? To test this, we compute the resolved molecular-to-stellar mass fraction, $rR^{\rm mol}_{\star}=\Sigma_{\rm mol}/\Sigma_{\star}$, for pixels within the bulge region from these two groups. The bottom panel of Figure \ref{fig_7B} shows the distribution of $rR^{\rm mol}_{\star}$ of pixels within bulges and adopting the fixed (hatched histograms) and variable (solid histograms) $\alpha_{\rm CO}$ prescriptions. On average, $rR^{\rm mol}_{\star}$ values of green valley pixels are $\sim3$ times lower than those within main sequence bulges when adopting the fixed $\alpha_{\rm CO}$. Although we note a displacement to the left of the mean $rR^{\rm mol}_{\star}$ values of the pixel distributions when adopting the variable $\alpha_{\rm CO}$, the tendencies do not change significantly ($rR^{\rm mol}_{\star}$ values for green valley galaxies are $\sim 5$ times lower than for main sequence galaxies). 

\begin{figure}
   \hspace{-0.5cm}
  \includegraphics[width=9cm]{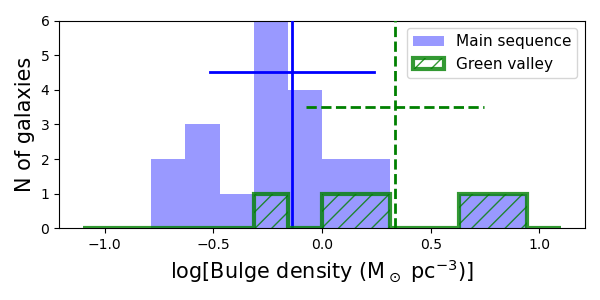}
  \caption{Distributions of the bulge density (in units of M$_\odot$ pc$^{-3}$), for the 23 main sequence and 5 green valley galaxies included in Figure \ref{fig_profiles}. The vertical and horizontal lines are the mean and the standard deviation values of the distributions, respectively. Although we do not see a statistically significant difference between green valley and main sequence bulge densities, we note that the former have on average denser bulges than the latter.}
  \label{fig_5}
\end{figure}

Similarly to the morphological quenching proposed by \cite{Martig2009}, numerical simulations performed by \cite{Gensior2020} show that bulges drive turbulence and increase the gas velocity dispersion, $\sigma_{\rm gas}$, virial parameter, and turbulent pressure, $P_{\rm turb}$, towards the galaxy centers. They note that the more compact and more massive (therefore, the more dense) the bulges are, the higher the level of turbulence. The star-formation activity is, therefore, ``dynamically suppressed'' in the innermost parts of bulge-dominant galaxies due to an increase of the gas turbulence that prevents the gravitational instabilities. Figure \ref{fig_5} shows the distribution of the bulge density, $\rho_{\rm b}$, for main sequence and green valley galaxies. To compute $\rho_{\rm b}$, we assume a spheroidal distribution of the bulge, i.e. we use $\rho_{\rm b}=M_{\rm b}/(\frac{4}{3}\pi R^3_{\rm b})$. Although  green valley galaxies are poorly represented, Figure \ref{fig_5} shows that, on average, green valley bulges tend to be more dense than those for main sequence galaxies. These results suggest that, when compared to main sequence pixels, the lower SFE$_{\rm mol}$ values within green valley bulges are not just a consequence of a poor molecular gas content. In addition, dynamical suppression may be reducing the star-formation rate in these regions due to an increase in $\Sigma_{\star}$ with $\Delta$SFMS (green valley bulges are $\sim 3$ times denser than those of main sequence galaxies). 

Which quenching mechanism is more important? In agreement with our results, recent studies support the idea that both changes in the gas reservoir and efficiency are responsible for reduced star formation in the disk of green valley galaxies. For instance, analyzing CO(1-0) data from the NOrthern Extended Millimeter Array (NOEMA) and ALMA for 7 nearby green valley galaxies, \cite{Brownson2020} show that the efficiency of star formation at their centers is on average three times lower than expected from the rKS (with some galaxies even up to 10 times less efficient). However, when they compare the resolved molecular gas main sequence (rMGMS, $\Sigma_\star$-$\Sigma_{\rm mol}$) and the rKS relations, they note that neither changes in the efficiency nor gas content dominate at $r\gtrsim0.6R_{\rm e}$. They conclude that while offsets from the rMGMS appear to dominate in the central regions, the full extents of the corresponding offsets from the rKS are unconstrained and make them unable to rank the two drivers in these regions. Similar results are shown by \cite{Lin2022}, who analyze the quenching mechanism in 22 green valley and 12 main sequence galaxies selected from ALMaQUEST. They note that the reduction of SFE and $R^{\rm mol}_{\star}$ in green valley galaxies (relative to main sequence galaxies) is seen in both bulge and disk regions (although with larger uncertainties). Their results thus suggest that, statistically, quenching in green valley galaxies may persist from the inner to the outer regions, and also that both gas depletion/removal and dynamical suppression are equally important.

}



\subsubsection{What drives star-formation quenching in ACA EDGE galaxies?}
\label{SF_quenching}

\begin{figure*}
  \includegraphics[width=18cm]{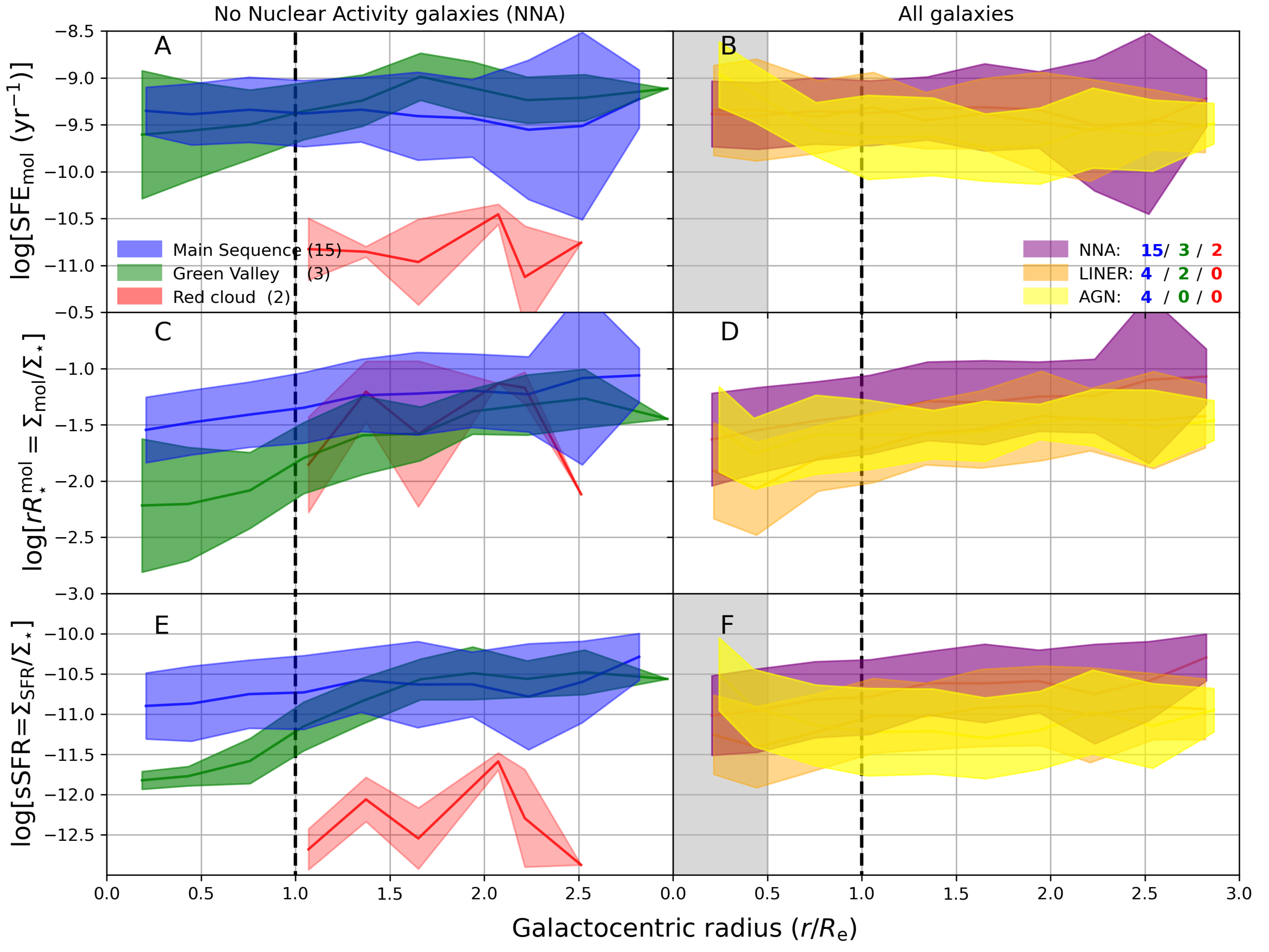} 
  \caption{The resolved star formation efficiency of the molecular gas, SFE$_{\rm mol}=\Sigma_{\rm SFR}/\Sigma_{\rm mol}$ (panels A and B), the resolved molecular-to-stellar mass fraction, $rR^{\rm mol}_{\star}=\Sigma_{\rm mol}/\Sigma_{\star}$ (panels C and D), and the specific star formation rate, sSFR$=\Sigma_{\rm SFR}/\Sigma_{\star}$ (panels E and F), in radial bins of 0.3$R_{\rm e}$ ($\sim$1.5 kpc) versus galactocentric radius for pixels from the 30 galaxies included in Figure \ref{fig_profiles}. The figure is color-coded according to the three main groups. Panels A, C, and E encompass pixels from 20 galaxies classified as SF (or with No Nuclear Activity, NNA; see column 5 in Table \ref{table_2}), split by their $\Delta$SFMS (i.e., main sequence, green valley, and red cloud) of the host galaxy. Panels B, D, and F include pixels from 30 ACA EDGE galaxies grouped according to the nuclear activity of the host galaxy. The grey shaded areas correspond to the regions where our H$\alpha$-based SFR estimator is susceptible to AGN/LINER contamination, so SFR and quantities related are only taken as upper-limits. In all panels, the vertical extent of the shaded areas is the $1\sigma$ scatter distribution for any group. Also, the vertical black dashed lines are located at $r= R_{\rm e}$, which we use to divide galaxy regions in central and disk pixels. While efficiencies in main sequence galaxies remain almost constant with galactocentric radius, in green valley galaxies we note a systematic increase of SFE$_{\rm mol}$, $rR^{\rm mol}_{\star}$, and sSFR, with increasing radius. We also observe slightly higher SFE$_{\rm mol}$ in the regions near the centers ($0.5 R_{\rm e} \lesssim r \lesssim 1.2 R_{\rm e}$) of AGNs when compared to their outskirts.}
  \label{fig_8}
\end{figure*}

The six panels in Figure \ref{fig_8} show the resolved SFE$_{\rm mol}$ (top panels), $rR^{\rm mol}_{\star}$ (middle panels), and the specific star formation rate, sSFR$=\Sigma_{\rm SFR}/\Sigma_{\star}$ (bottom panels), versus galactocentric radius (in radial bins of $0.3 R_{\rm e}$, $\sim 1.5$ kpc resolution at the mean distance) for the 30 galaxies included in Figure \ref{fig_profiles} (i.e. the 30 ACA EDGE galaxies with $i<70^{\circ}$ and $5\sigma$ CO detections). In order to better understand the different mechanisms behind star-formation quenching in ACA EDGE galaxies, we split the panels of Figure \ref{fig_8} into two groups. Panels A, C, and E include SF galaxies (hereafter no nuclear activity galaxies, NNA; i.e., pixels from galaxies without LINER/AGN activity), split by their $\Delta$SFMS (i.e., main sequence, green valley, and red cloud). Panels B, D, and F include pixels from NNA, LINER, and AGN galaxies (shaded purple, orange, and yellow regions, respectively), according to their nuclear activity (column 5 in Table \ref{table_2}). 

On average, the SFE$_{\rm mol}$ remains almost constant with radius for NNA main sequence, green valley, and red cloud galaxies (panel A). These results are consistent with \cite{Villanueva2021}; while they do not observe significant variations of SFE$_{\rm mol}$ with radius in the CARMA EDGE sample, they also note a systematic decrease in the efficiencies from late- to early-type galaxies. In addition, panel A shows that green valley galaxies have a mild increase in SFE$_{\rm mol}$ with radius. While main sequences and green valley galaxies have similar $rR^{\rm mol}_{\star}$ for $r\gtrsim 1.8 R_{\rm e}$ (see panel C), the latter have significantly lower $rR^{\rm mol}_{\star}$ than the former at $r\lesssim 1.5 R_{\rm e}$. molecular-to-stellar mass fractions for green valley galaxies can reach values even $\sim$0.8 dex below than those for main sequence at $r\lesssim 0.5 R_{\rm e}$. Similarly, sSFRs show almost the same radial trends as that of $rR^{\rm mol}_{\star}$ (see panel E). The sSFR values in green valley galaxies are typically about an order of magnitude below those of main sequence galaxies ($\sim$1.2 dex). These results suggest that what is driving the star-formation quenching in green valley galaxies is related to both a decrease of the SFR (e.g., via changes in the star-formation efficiency) and gas removal and/or depletion.

\begin{figure*}
\hspace{-1.5cm}
  \includegraphics[width=21.5cm]{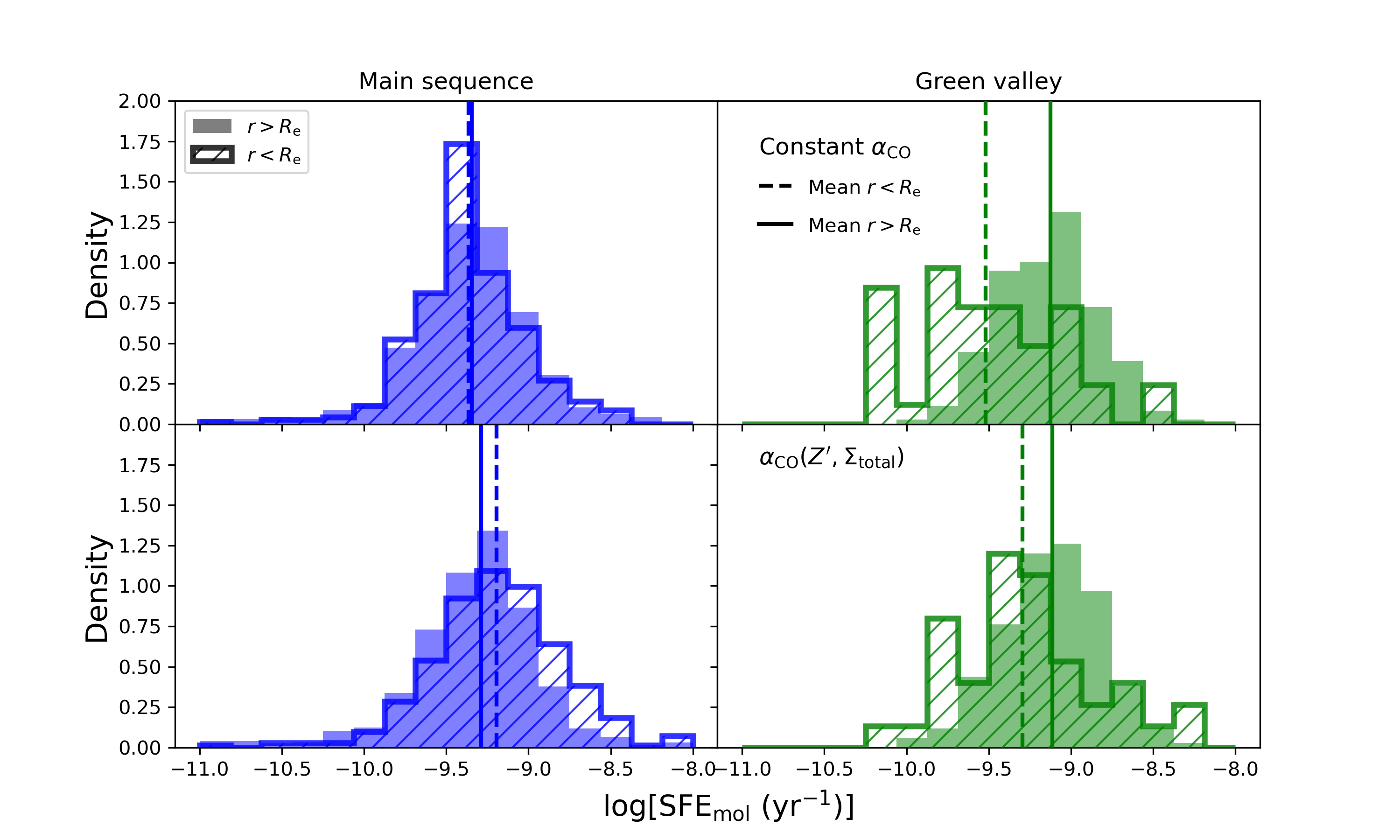} 
  \caption{SFE$_{\rm mol}$ distributions for pixels from no nuclear activity galaxies (NNA), split in main sequence (blue bars) and green valley (green bars) galaxies (from left to right panels, respectively). The two groups are split by two radial bins according to the breaks identified in Fig. \ref{fig_8}, thus between pixels within the central (hatched unfilled bars) and outer (solid bars) regions. To compute the SFE$_{\rm mol}$, we adopt a fixed CO-to-H$_2$ conversion factor (top panels), and the variable $\alpha_{\rm CO}(Z',\Sigma_{\rm total})$ from Equation \ref{eq:alpha_co} (bottom panels). While the distributions of SFE$_{\rm mol}$ for main sequences pixels within the two radial bins are similar when adopting the two $\alpha_{\rm CO}$ prescriptions, green valleys show a more clear bimodal behaviour when using a constant $\alpha_{\rm CO}$.}
  \label{fig_9A}
\end{figure*}

Similar to panel A of Figure \ref{fig_8}, panel B shows that NNA galaxies (mostly dominated by the main sequence) have on average flat SFE$_{\rm mol}$ profiles. Although both LINER and AGN galaxies have remarkably high efficiencies in the central regions ($r \lesssim 0.5 R_{\rm e}$; grey shaded area in panels B and F), these values have to be considered carefully due to LINER/AGN contamination (as explained in \S \ref{SFE_bulge}). Consequently, SFE$_{\rm mol}$ (and quantities related) must be considered only as upper-limits for these two groups within this region. While LINERs and SFs show a flat SFE$_{\rm mol}$ profile for $r \gtrsim 0.5 R_{\rm e}$, AGNs seem to have significantly lower efficiencies in the range $ 0.75 R_{\rm e} \lesssim r \lesssim 2.0 R_{\rm e}$ than LINER/NNA galaxies, which finally flatten at larger galactocentric radii. When analyzing $rR^{\rm mol}_{\star}$ as a function of galactocentric radius (shown in panel D), we observe a systematic inside-out increase of the molecular fractions for with radius for each of the three groups. However, LINERs/AGNs have $rR^{\rm mol}_{\star}$ values slightly lower than NNA galaxies ($\sim 0.2-0.5$ dex below) for the galactocentric radius range covered here. 
\noindent We also note that, on average, sSFR has a similar behaviour as SFE$_{\rm mol}$, particularly for AGN galaxies which show a slight decrease of the sSFR with radius (similar to the one seen for SFE$_{\rm mol}$). This may be suggesting that AGN activity mitigates the star formation activity, although not necessarily by impacting the H$_2$ reservoirs (e.g., \citealt{Bluck2020,Bluck2020b}).

Our results are consistent with CALIFA-based studies reporting lower molecular gas fractions in centers of AGN hosting galaxies when compared to their outskirts (e.g., \citealt{Sanchez2018, Lacerda2020,Ellison2021}). However, observational evidence has also shown that the gas content in AGN hosts can be similar (or even higher) than galaxies without nuclear activity, either by analyzing the atomic (e.g., \citealt{Ho2008,Fabello2011,Ellison2019}), or molecular (e.g., \citealt{Maiolino1997,Saintonge2017,Koss2021,Esposito2022}) gas reservoirs.

\begin{figure*}
\hspace{-1.5cm}
  \includegraphics[width=21.5cm]{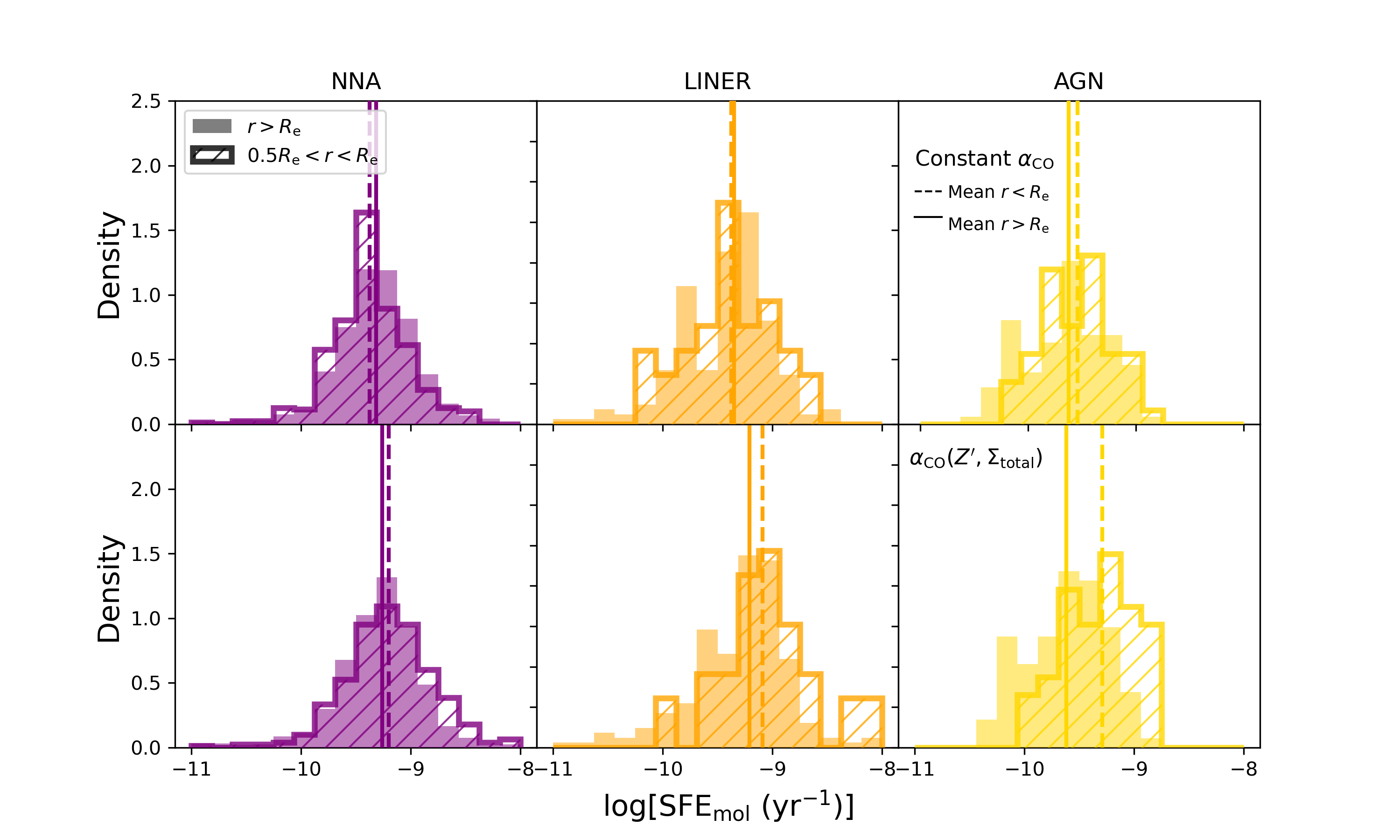} 
  \caption{SFE$_{\rm mol}$ distributions for pixels from star-forming (i.e., galaxies with No Nuclear Activity, NNA; purple bars), LINER (orange bars), and AGN (yellow bars) galaxies (from left to right panels, respectively). Conventions are as in Fig. \ref{fig_9A}. While NNA and LINER pixels have similar SFE$_{\rm mol}$ distributions for the two radial bins and when testing the two $\alpha_{\rm CO}$ prescriptions, we note a mild bimodal behaviour for AGNs.}
  \label{fig_9B}
\end{figure*}

These results suggest that the cessation of the star-formation activity has different modes depending on galaxy substructures, morphological type, and nuclear activity. NNA main sequence and green valley galaxies have SFE$_{\rm mol}$ consistent with local star-forming spirals \citep[e.g.,][]{Villanueva2021,Villanueva2022}, which on average remain constant with radius. Nevertheless, green valley galaxies show signs of an inside-out increase in their efficiencies. {To better understand these differences, we compute the SFE distributions for NNA galaxies by splitting them in to central pixels (i.e., pixels at $r<R_{\rm e}$) and outer pixels (i.e., pixels at $r>R_{\rm e}$). 
\noindent We also test how these distributions change with the two $\alpha_{\rm CO}$ prescriptions included in this work (as shown in Figure \ref{fig_9A})}. The distribution of SFE$_{\rm mol}$ for main sequence galaxies is almost identical when we split their pixels in to two radial bins at $r = R_{\rm e}$. If we adopt fixed $\alpha_{\rm CO}$ (top panels), green valley and red cloud pixels show a clear bimodal behaviour. We test how the SFE distributions change by using the variable $\alpha_{\rm CO}(Z',\Sigma_{\rm total})$ prescription (bottom panels). Interestingly, we note that  green valley galaxies show mild bimodal distributions. 
\noindent We perform a Student's $t$-test to verify if the distribution of SFE$_{\rm mol}$ values in green valley galaxies are drawn from the same parent population. We obtain $\lvert t \rvert =0.89$ for green valley (degrees of freedom = $222$) pixel distributions, which is below the critical $t$-value $t_{\alpha=0.05}\approx1.96$; we thus can reject the null hypothesis that the two green valley groups are drawn from the same underlying distribution with 95\% confidence. Although these results suggest that morphological quenching may be acting after the gas removal stage in green valley galaxies \citep[e.g.,][]{Colombo2020}, the small difference between these two distributions may be caused by the poor spatial resolution of our CO observations ($\sim$1.5 kpc) when compared to the physical scale required to resolve bulges in ACA EDGE galaxies ($\lesssim 500$ pc). In addition, some studies \citep[e.g.,][]{Cook2019,Cook2020} have discarded a scenario where bulges play a key role in controlling the star-formation activity, suggesting that this could be reflecting physical processes more associated with galaxy disks. Finally, when analyzing the individual SFE$_{\rm mol}$ pixel distributions within $R_{\rm e}$ for the three green valley galaxies included in left panels of Figure \ref{fig_9A} and using the morphological and bar classification included in \cite{Kalinova2021} for CALIFA galaxies, we note that spiral galaxies without bars (i.e., NGC 7716) seem to have higher efficiencies than those with a predominant bar on their disks (i.e., UGC 12250 and NGC 0171). However, due to the limited galaxy sample included in this analysis, it is essential that future ACA EDGE survey studies increase the green valley coverage to derive more statistically significant conclusions about how structural components (e.g., bars) could enhance the effects of morphological quenching.

Similarly to Figure \ref{fig_9A}, Figure \ref{fig_9B} includes the SFE$_{\rm mol}$ distributions for two radial bins, i.e. for pixels within $r<1.2R_{\rm e}$ (hatched histograms) and at $r>1.2R_{\rm e}$ (solid histograms), in NNA (purple bars), LINER (orange bars), and AGN (yellow bars) galaxies. We also test how the distributions change with the two $\alpha_{\rm CO}$ prescriptions. To avoid SFR contamination due to AGN/LINER, we reject pixels at $r<0.5 R_{\rm e}$. While NNA, LINER, and AGN pixels have similar distributions for the two radial bins and using the fixed $\alpha_{\rm CO}$ (top panels of Fig. \ref{fig_9B}), we note signs of a bimodal behaviour for AGNs if we adopt the variable $\alpha_{\rm CO}(Z',\Sigma_{\rm total})$ prescription (bottom left panel). 
\noindent We perform a Student's $t$-test to verify if the AGN distributions are drawn from the same parent population; we obtain $\lvert t \rvert =1.89$ (degrees of freedom = $140$), which is lower than the critical $t$-value $t_{\alpha=0.05}\approx1.97$. We thus can reject the null hypothesis that the two AGN groups are drawn from the same underlying distribution. Although SFE$_{\rm mol}$ values for AGNs are consistent with observational evidence showing that optical and radio selected AGNs tend to have similar/lower SFRs than typical main sequence galaxies (e.g., \citealt{Ellison2016,Sanchez2018,Lacerda2020}), which appears to {be mainly due in SFR within galaxy centers} (e.g., \citealt{Ellison2018,Sanchez2018,Kalinova2021}), these results could be also supporting the idea of a slight enhancement of the star formation in these regions. {However, studies have shown that the impact of AGN ionization can reach as far out as 10s of kpc \citep[e.g.,][]{Veilleux2003,Husemann2008,Nesvadba2011}. Although unlikely, we cannot rule out that the high SFEs we measure at the centers of ACA EDGE AGNs are due to contamination by AGN emission, even though we have excluded pixels with $r<0.5R_{\rm e}$. }

Morphological quenching has been shown to be a good candidate to explain the decrease of the SFE$_{\rm mol}$ observed in green valley ACA EDGE galaxies, perhaps via gas stabilization or dynamical suppression \citep[e.g.,][]{Martig2009,Gensior2020,Gensior2021}, increasing the turbulent velocity dispersion of the gas (e.g., \citealt{Vollmer2013}), due to a sequence of short-lived AGN (e.g., \citealt{Bluck2020,Bluck2020b}), or a combination of mechanisms (e.g., \citealt{Lin2019}). However, the similarity of the SFE distributions shown in the bottom panels of Figures \ref{fig_9A} and \ref{fig_9B} (particularly for green valley galaxies) {suggest that these processes have a minimal impact on the efficiencies}. These mechanisms seem to respond to non-long-standing processes and may only complement the gas depletion and/or removal. In addition, recent studies have shown that the presence of a classical bulge seems to not be the only necessary condition for morphological quenching in nearby galaxies. For example, \cite{Kalinova2022} find that some galaxies with large central bulges may actually correspond to star-forming systems, and conversely some galaxies with small spheroids may be quenched. They also note that higher central surface densities ($\sim 10^{4}$ M$_\odot$ pc$^{-2}$), no bars, and early-type morphologies (i.e., no tight and prominent spiral arms) seem to be either connected or an additional condition for dynamical suppression in galaxies. 

Further studies based on CO data within galaxy centers with both higher resolution and sensitivity than those presented in this work (e.g., at physical scales $\lesssim 500$ pc) could give us more information about the dynamical state of the molecular gas within bulges of green valley and red cloud galaxies. These are essential to disentangle the actual connection between the SFE$_{\rm mol}$ and the gravitational stability of the gas, or the effects of AGN in the star-formation activity in detail.

\section{Summary and conclusions}
\label{S5_Conclusions}

We present a systematic study of the star formation efficiency and its dependence on other physical parameters in 60 galaxies from the ACA EDGE survey. We analyze $^{12}$CO($J=2-1$) data cubes and optical IFU data from CALIFA. {Compared to other local galaxy surveys, ACA EDGE is designed to mitigate selection effects based on CO brightness and morphological type. This results in a less biased galaxy survey and an ideal sample to investigate the effects of star-formation quenching on massive local galaxies.} We conduct a detailed analysis to characterize the main properties of the molecular gas by deriving global (e.g., integrated masses and SFRs) and resolved quantities out to typical galactocentric radii of $r\approx 3 R_{\rm e}$. We use a constant Milky Way CO-to-H$_2$ conversion factor $\alpha_{\rm CO,MW}=4.3$ M$_\odot$ $(\rm K \, km \, s^{-1} \, pc^{2})^{-1}$ \citep[][]{Walter2008} and a Rayleigh-Jeans brightness temperature line ratio of $R_{21}=I_{\rm CO(2-1)}/I_{\rm CO(1-0)}\sim 0.65$. We also test the impact of the constant CO-to-H$_2$ conversion factor adopted in our results by using the variable $\alpha_{\rm CO}(Z',\Sigma_{\rm total})$ from \cite{Bolatto2013}. We conduct a systematic analysis to explore molecular and stellar scale lengths, bulge physical properties, molecular-to-stellar mass fractions, and the SFE of the molecular gas in ACA EDGE galaxies to compare them with the current literature. Our main conclusions are enumerated as follows:

\begin{enumerate}
    \item We compute the molecular depletion times, $\tau_{\rm dep}$, of ACA EDGE galaxies. Although the majority of galaxies have $\tau_{\rm dep}\sim1$ Gyr, we find that molecular depletion times varies significantly with distance of the SFR to the star formation main sequence line, $\Delta$SFMS. Classifying galaxies as main sequence ($-0.5$ dex$\leq \Delta$SFMS$\leq0.5$ dex), green valley ($-1.0$ dex$< \Delta$SFMS$\leq-0.5$ dex), and red cloud ($\Delta$SFMS$\leq-1.0$ dex) galaxies, we note a systematic decrease in the molecular-to-stellar mass fraction, $R^{\rm mol}_{\star}$, and an increase in $\tau_{\rm dep}$ with $\Delta$SFMS (see Fig. \ref{fig_3}).

    \item We determine the molecular and stellar exponential disk scale lengths, $l_{\rm mol}$ and $l_{\star}$, respectively (see Fig. \ref{fig_4}). We fit an exponential function to 23 molecular gas surface density, $\Sigma_{\rm mol}$, and 30 stellar surface density, $\Sigma_{\star}$, radial profiles from  the 30 ACA EDGE galaxies with $5\sigma$ CO detections and inclinations $<70^\circ$. We find a close 6:5 relation between $l_{\rm mol}$ and $l_{\star}$ ($l_{\star}=[1.24\pm0.05]\times l_{\rm mol}$), which is consistent with previous results from the literature for main sequence spirals (e.g., HERACLES, CARMA EDGE).

    \item We derive the $\Sigma_{\rm SFR}$-$\Sigma_{\star}$ and $\Sigma_{\rm SFR}$-$\Sigma_{\rm mol}$ relations, the resolved star formation main sequence (rSFMS) and the resolved Kennicutt-Schmidt (rKS) relations, respectively (see Fig. \ref{fig_6}). We find slopes of $\alpha_{\rm rSFMS}=[1.20\pm0.07]$ and $\alpha_{\rm rKS}=[1.19\pm0.07]$ for the rSFMS and rKS. Although the slopes for ACA EDGE galaxies are larger than those of spiral star-forming main-sequence galaxies selected from the field (e.g., CARMA EDGE, PHANGS), they are consistent with those found in galaxy surveys that are more oriented to increase the coverage of green valley and red cloud galaxies (e.g., ALMaQUEST). However, we remark that these slopes are very sensitive to the fitting method and the $\alpha_{\rm CO}$ prescription adopted.

    \item 
    \noindent We compute the resolved star-formation efficiency of the molecular gas, SFE$_{\rm mol}$, within the bulge region of 23 main sequence and 5 green valley ACA EDGE galaxies. We find that SFE$_{\rm mol}$ values within green valley bulges tend to be lower than for main sequence galaxies ($\sim3$ times lower). The results suggest that in addition to poor molecular gas content, dynamical suppression may be reducing the star-formation rate in the bulge region of green valley galaxies due to a decrease in SFE$_{\rm mol}$ with $\Delta$SFMS (see Fig. \ref{fig_7B}).

    \item We compute radial profiles for SFE$_{\rm mol}$, the resolved molecular-to-stellar mass fraction $rR^{\rm mol}_{\star}=\Sigma_{\rm mol}/\Sigma_{\star}$, and the resolved specific star formation rate sSFR$=\Sigma_{\rm SFR}/\Sigma_{\star}$, for pixels grouped according to their $\Delta$SFMS and nuclear activity (see Fig. \ref{fig_8}). We note a systematic decrease in SFE$_{\rm mol}$, $rR^{\rm mol}_{\star}$, and sSFR with $\Delta$SFMS. We also observe a slight inside-out increase in the efficiencies in green valley galaxies out to $r\approx R_{\rm e}$; from this point on, SFE$_{\rm mol}$ increases until it reaches similar values to the almost constant values we observe for main sequence galaxies. 
    \noindent Although the efficiencies of green valley galaxy centers are more similar to those of their outer disks when we use the variable $\alpha_{\rm CO}(Z',\Sigma_{\rm total})$ prescription, on average their SFE$_{\rm mol}$ distributions show lower efficiencies in their central regions when compared to both those for their outskirts ($\sim2$-3 times lower) and the typical values of main-sequence galaxies ($\sim2$ times lower; see Fig. \ref{fig_9A}).  
    
\end{enumerate}

Our results suggest that although gas depletion and/or removal seem to be the most important mechanisms behind the cessation of stellar production, they do not completely explain the star-formation quenching processes in ACA EDGE galaxies. Complementary mechanisms (such as morphological quenching and/or AGN feedback) are therefore required to change the physical properties of the molecular gas, which could impact its ability to form stars in galaxies transiting through the green valley. The inside-out nature of these processes is reflected by the decreasing of the SFE$_{\rm mol}$ in the central regions of green valley galaxies, although this change is dependant on the $\alpha_{\rm CO}$ prescription adopted. Future projects should focus on increasing the early-type galaxy coverage to improve the statistical significance of these results. In addition, high resolution CO observations in the central parts of green valley and red cloud galaxies are essential to better understand how these mechanisms may impact the stability of the gas at physical scales comparable to those of molecular clouds ($\lesssim 100$ pc).

\begin{acknowledgments}

V. V. acknowledges support from the scholarship ANID-FULBRIGHT BIO 2016 - 56160020, funding from NRAO Student Observing Support (SOS) - SOSPADA-015, and funding from the ALMA-ANID Postdoctoral Fellowship under the award ASTRO21-0062. A. D. B, S. V., and V. V., acknowledge partial support from NSF-AST1615960. R.C.L. acknowledges support for this work provided by a National Science Foundation (NSF) Astronomy and Astrophysics Postdoctoral Fellowship under award AST-2102625. This work is also supported by NSF under the awards AST-2307440 and AST-2307441.

\end{acknowledgments}

\software{ Astropy \citep{AstropyCollaboration2018}, MatPlotLib \citep{Hunter2007},
NumPy \citep{Harris2020}, SciPy \citep{2020SciPy-NMeth}, seaborn \citep{Waskom2021}, Scikit-learn \citep{scikit-learn}.}

\appendix
\label{appendix}
\subsection{Multi-panel Images}

\begin{figure*}
\hspace{.5cm}
  \includegraphics[width=16.cm]{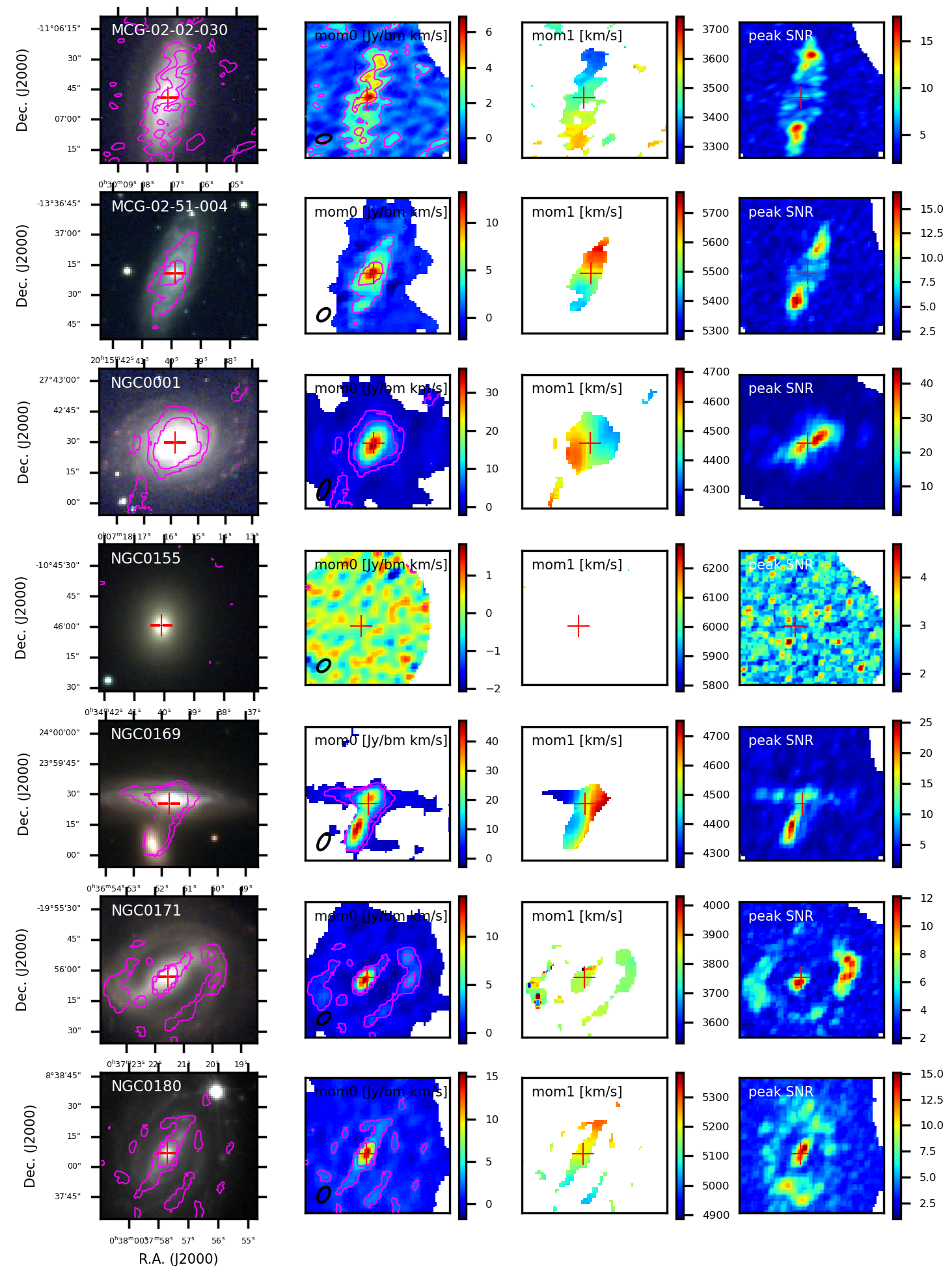}
  \caption{Images for ACA EDGE galaxies. See caption in Figure \ref{fig_mom0_1}.}
  \label{fig_mom0_2}
\end{figure*}

\begin{figure*}
\hspace{.5cm}
  \includegraphics[width=16.cm]{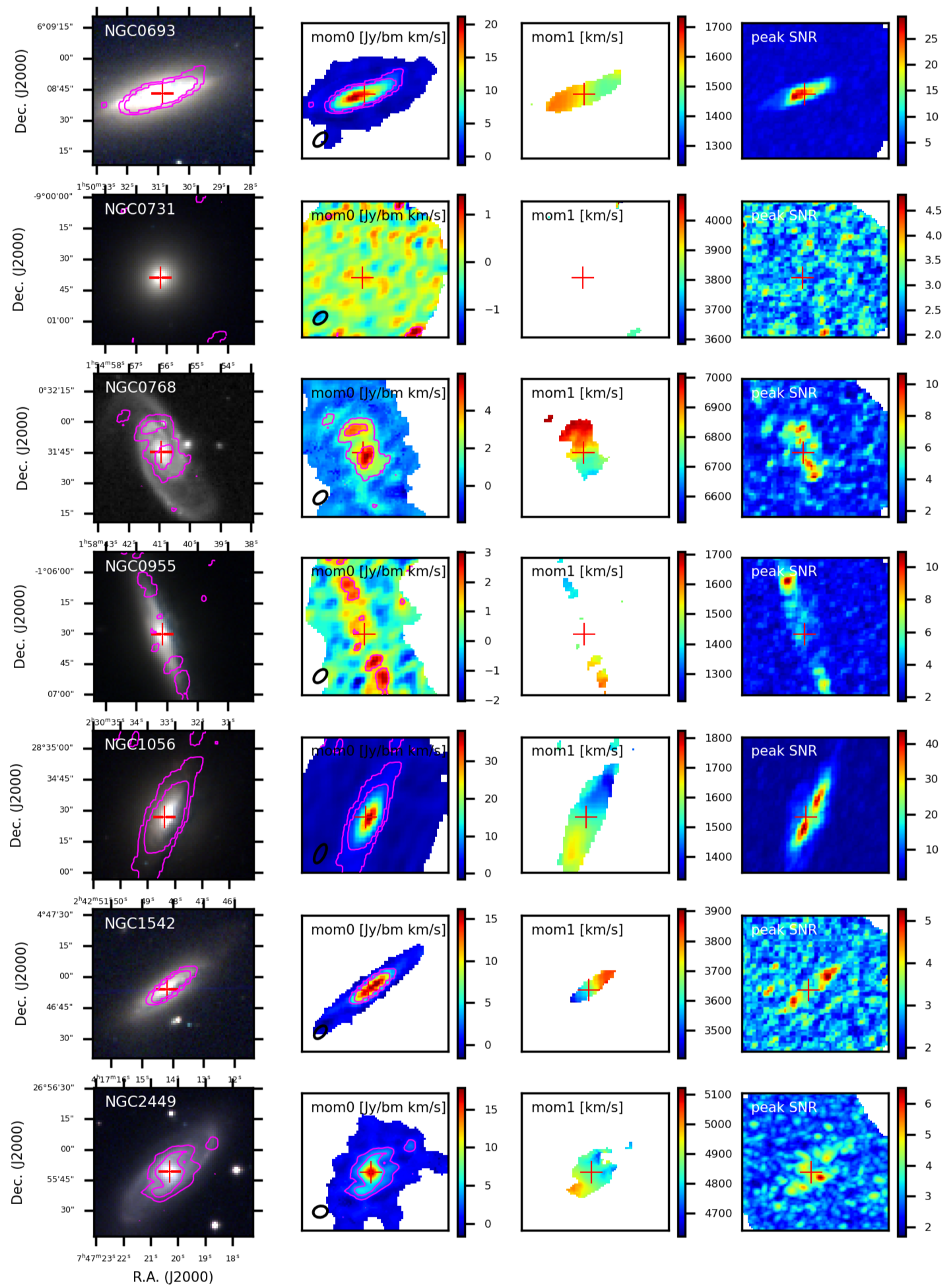}
  \caption{Images for ACA EDGE galaxies. See caption in Figure \ref{fig_mom0_1}.}
  \label{fig_mom0_3}
\end{figure*}

\begin{figure*}
\hspace{.5cm}
  \includegraphics[width=16.cm]{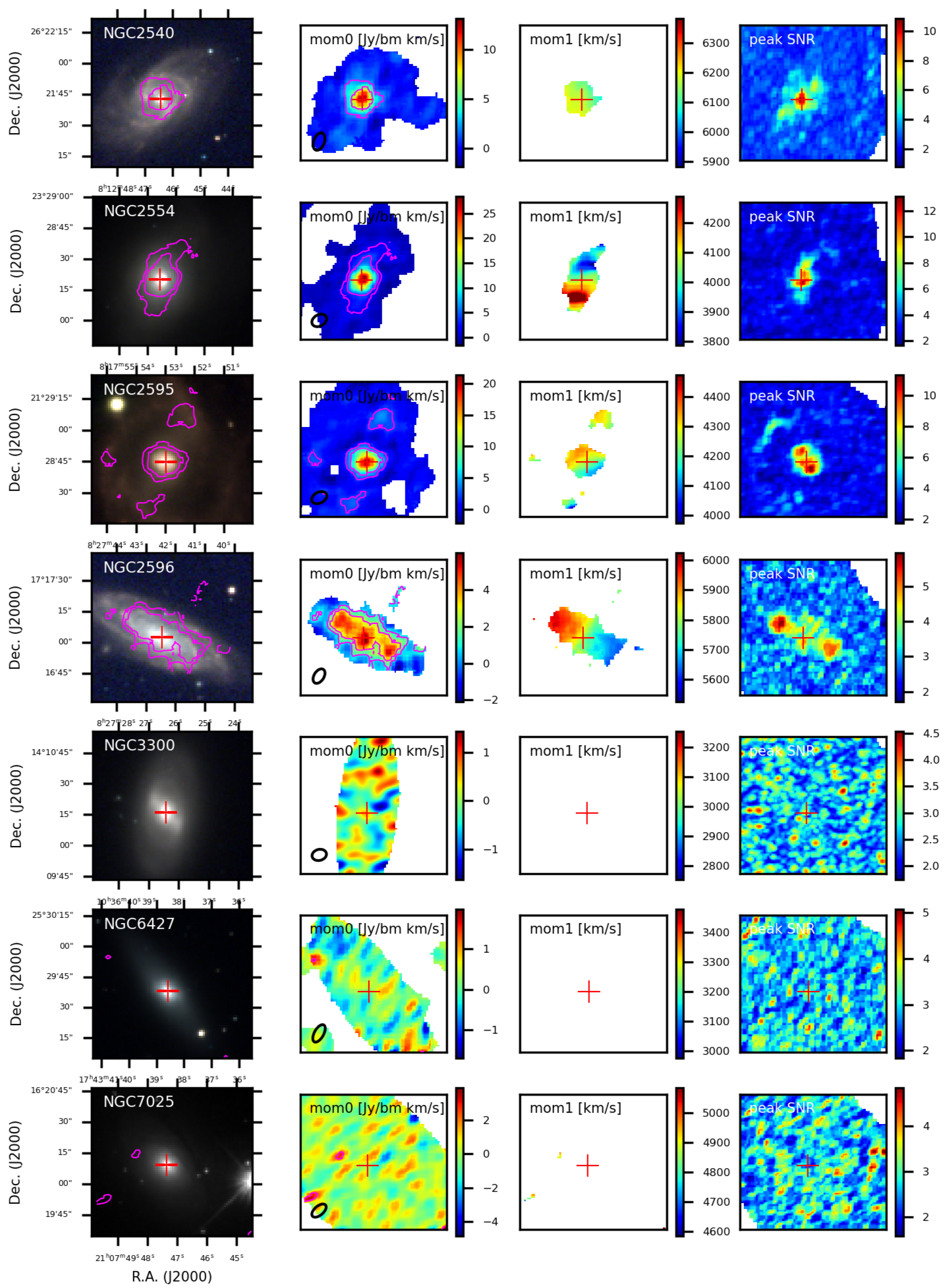}
  \caption{Images for ACA EDGE galaxies. See caption in Figure \ref{fig_mom0_1}.}
  \label{fig_mom0_4}
\end{figure*}

\begin{figure*}
\hspace{.5cm}
  \includegraphics[width=16.cm]{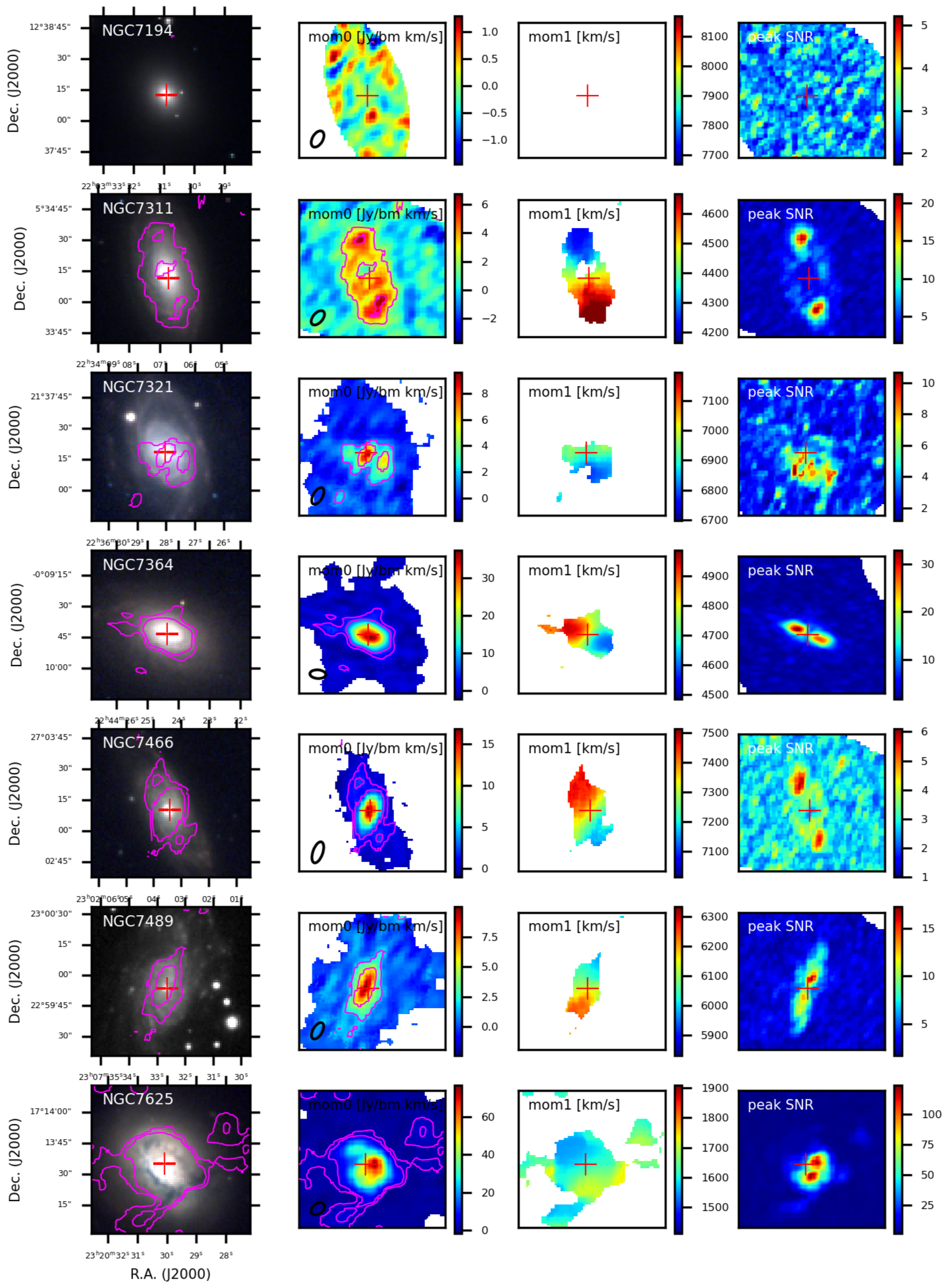}
  \caption{Images for ACA EDGE galaxies. See caption in Figure \ref{fig_mom0_1}.}
  \label{fig_mom0_5}
\end{figure*}

\begin{figure*}
\hspace{.5cm}
  \includegraphics[width=16.cm]{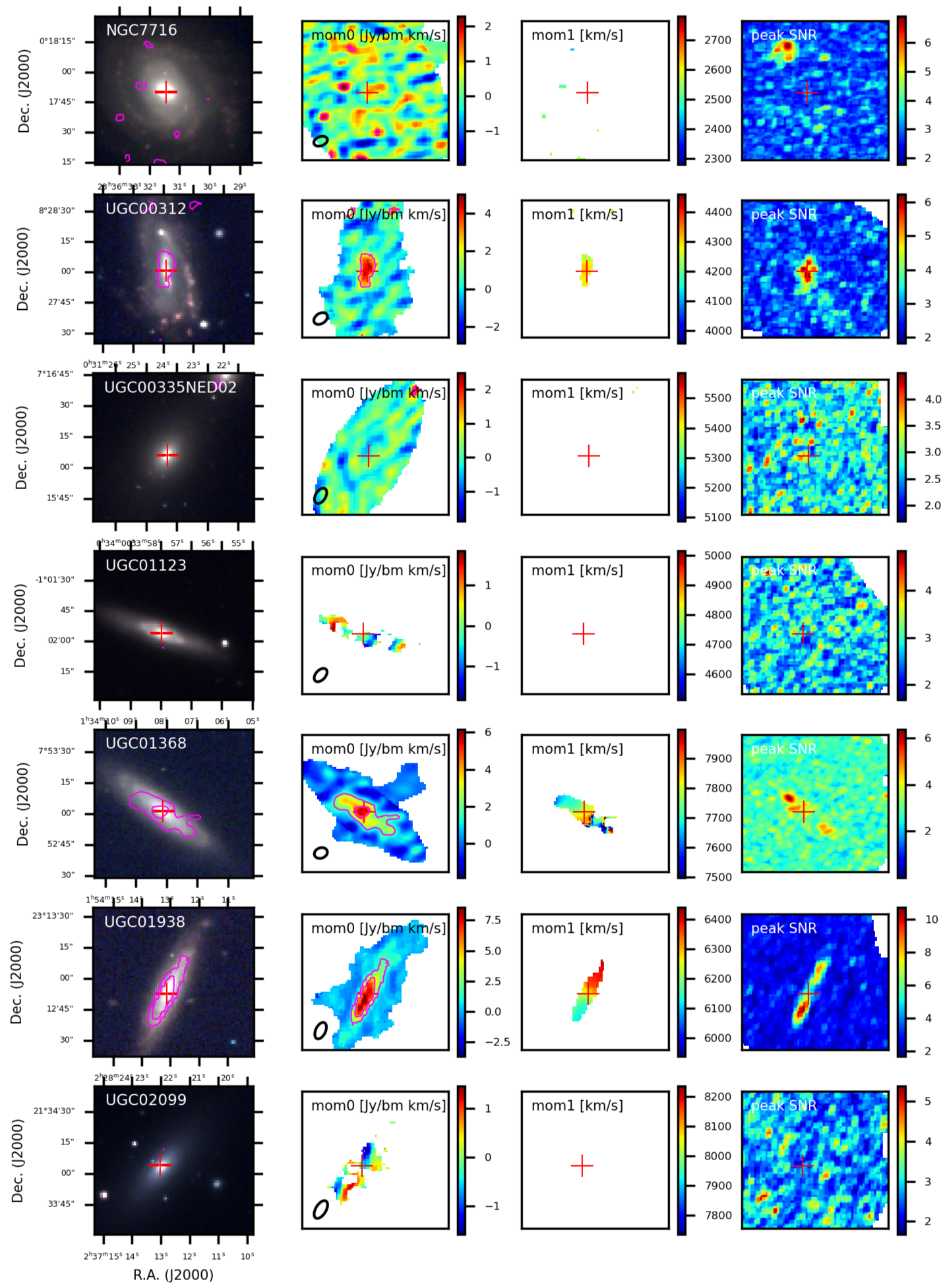}
  \caption{Images for ACA EDGE galaxies. See caption in Figure \ref{fig_mom0_1}.}
  \label{fig_mom0_6}
\end{figure*}

\begin{figure*}
\hspace{.5cm}
  \includegraphics[width=16.cm]{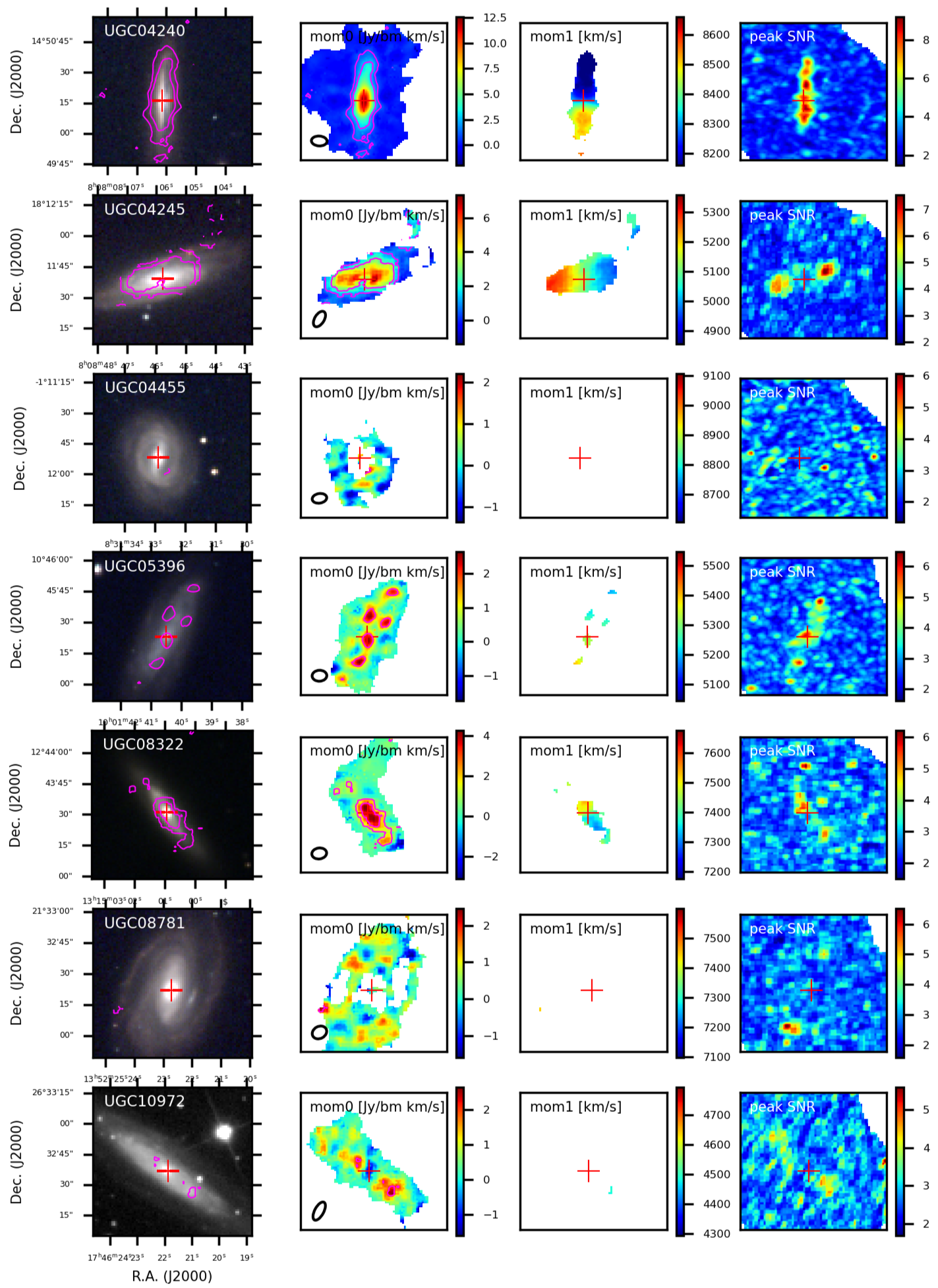}
  \caption{Images for ACA EDGE galaxies. See caption in Figure \ref{fig_mom0_1}.}
  \label{fig_mom0_7}
\end{figure*}

\begin{figure*}
\hspace{.5cm}
  \includegraphics[width=16.cm]{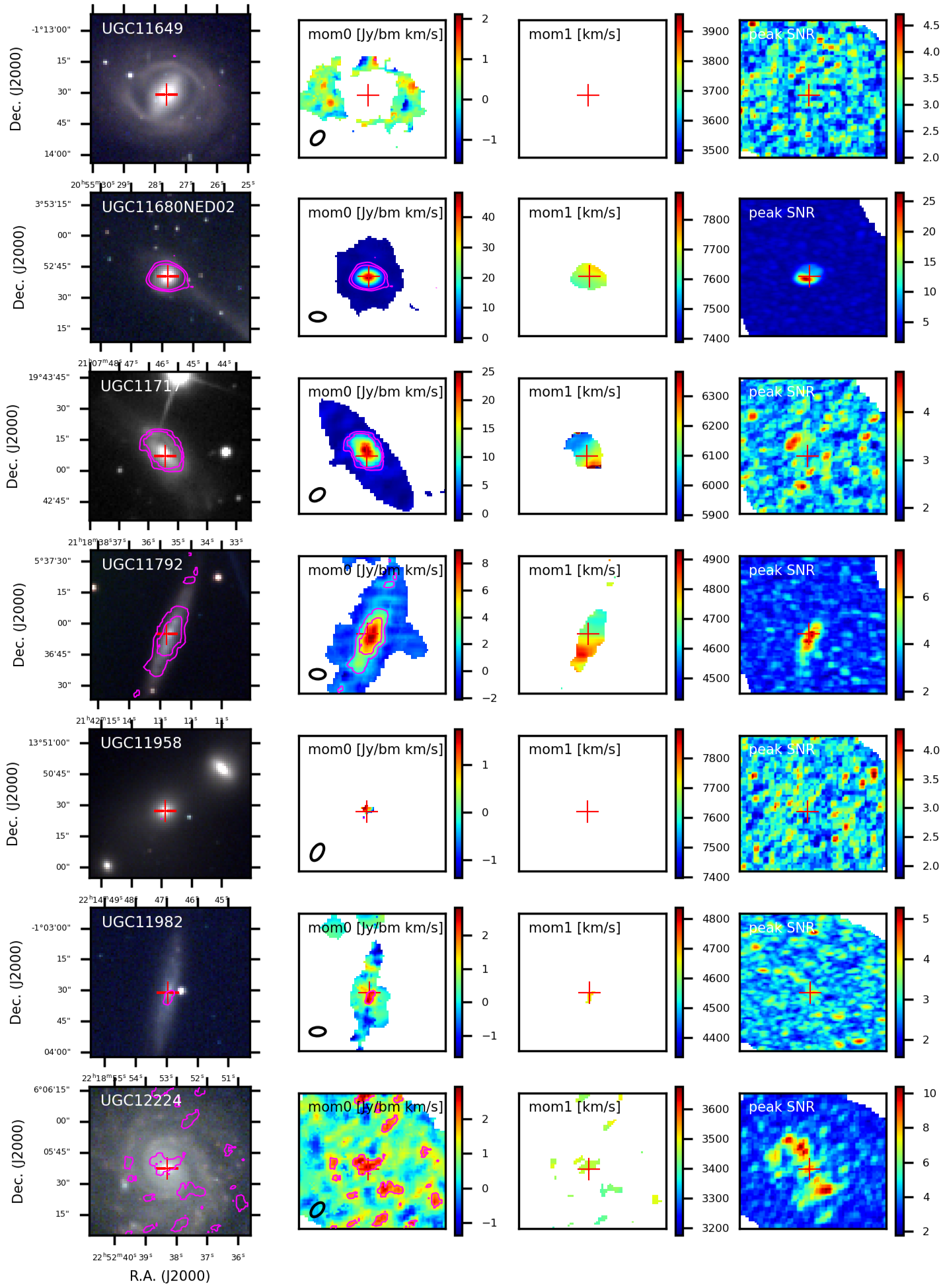}
  \caption{Images for ACA EDGE galaxies. See caption in Figure \ref{fig_mom0_1}.}
  \label{fig_mom0_8}
\end{figure*}

\begin{figure*}
\hspace{.5cm}
  \includegraphics[width=16.cm]{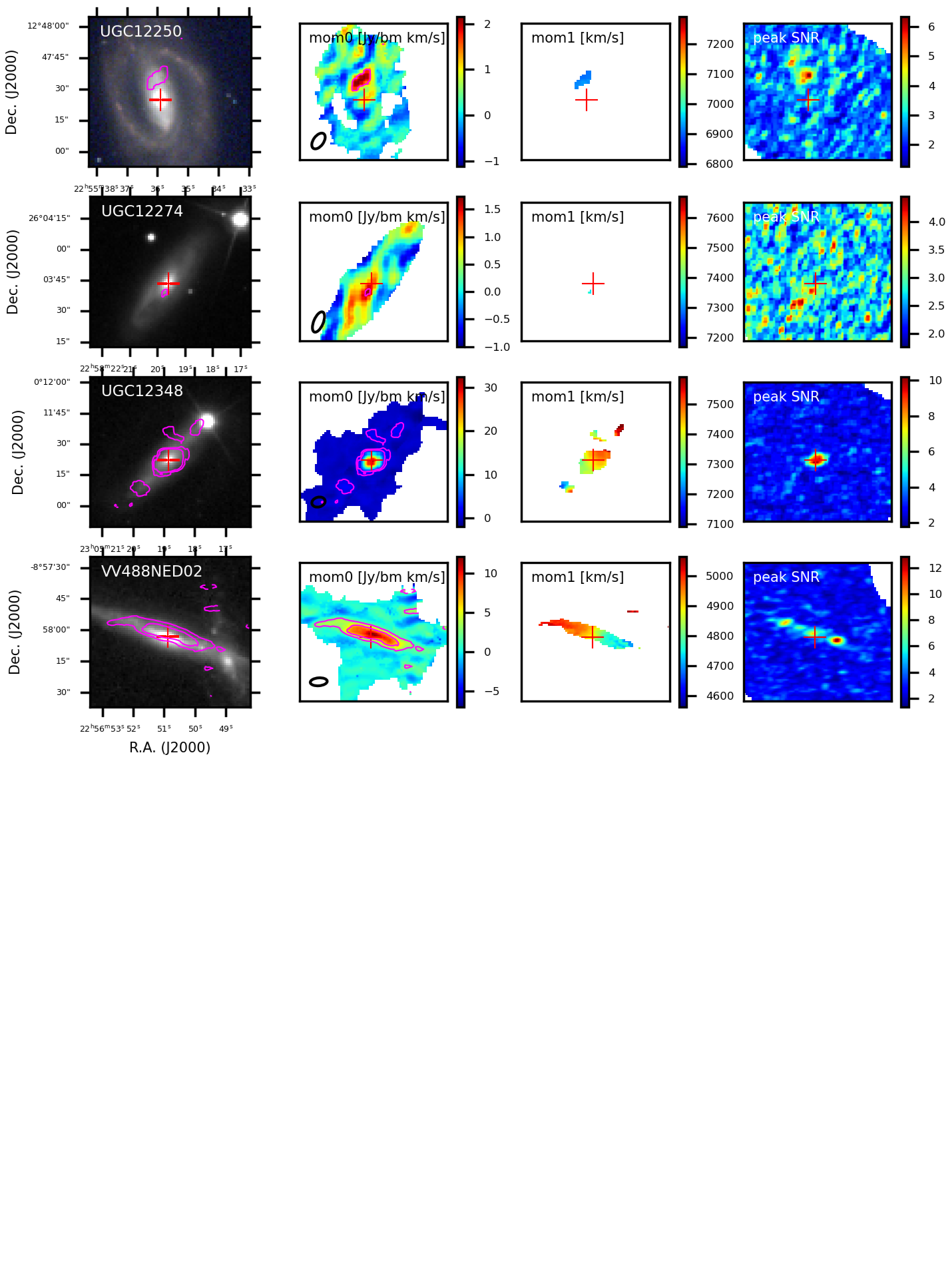}
  \caption{Images for ACA EDGE galaxies. See caption in Figure \ref{fig_mom0_1}.}
  \label{fig_mom0_9}
\end{figure*}

Figures \ref{fig_mom0_2} to \ref{fig_mom0_9} in this appendix follow the same format as Figure \ref{fig_mom0_1}, and show the products for 53 galaxies included in the ACA EDGE survey.


\bibliography{main}{}
\bibliographystyle{aasjournal}



\end{document}